\documentclass[a4paper,fleqn,usenatbib]{mnras}

\usepackage{newtxtext,newtxmath}

\usepackage[T1]{fontenc}
\usepackage{ae,aecompl}
\usepackage{graphicx}	
\usepackage{amsmath}	
\usepackage{longtable}
\usepackage{pdflscape}
\usepackage{natbib}
\usepackage{textcomp}
\newcommand{\teff}{$T_\textrm{eff}$}
\newcommand{\logg}{$\log g$}

\newcommand{\kms}{km\,s$^{-1}$}
\newcommand{\vsini}{$v\sin i$}
\newcommand{\vrad}{$v_{\rm rad}$}

\newcommand{\tess}{{\it TESS}}

\newcommand{\ktwo}{{\it K2}}
\newcommand{\plato}{{\it PLATO}}

\title[\ktwo/\tess–HERMES study of bright field stars]{High-precision \ktwo/\tess\ Photometry and HERMES Spectroscopy of Four Bright Field Stars}

 \author[Trust et al.]{
 {Otto Trust}$^{1,2,3}$\thanks{E-mail: otrust@must.ac.ug},
 {Santosh Joshi}$^4$,
 {Benard Nsamba}$^{2,3,5}$, 
 {Sharon Aol}$^1$,
 {Owen Vermeulen}$^6$,
 \newauthor
 {Ronald Ssembatya}$^2$,
 {Peter De Cat}$^{7}$,
 {Patricia Lampens}$^{7}$,
 {Sydney A. Barnes}$^8$,
 {Partha P. Goswami}$^9$,
 \newauthor
 {Athul Dileep}$^{4}$,
 {Surath C. Ghosh}$^{4,10}$,
 { Pramod S. Kumar}$^{11}$,
{Edward Jurua}$^{1}$,
{Mrinmoy Sarkar}$^{4}$,
\newauthor
{Sarabjeet S. Bedi}$^{10}$\\
\\
 $^{1}$Department of Physics, Mbarara University of Science and Technology, P.O. Box 1410, Mbarara, Uganda\\
 $^{2}$Department of Physics, Faculty of Science, Kyambogo University, P.O. Box 1, Kyambogo, Uganda\\
 $^{3}$Max-Planck-Institut für Sonnensystemforschung, Justus-von-Liebig-Weg 3, 37077 Göttingen, Germany\\
 $^{4}$Aryabhatta Research Institute of Observational Sciences, Manora Peak, Nainital- 263002, India\\
 $^{5}$Max-Planck-Institut für Astrophysik, Karl-Schwarzschild-Str. 1, D-85748 Garching, Germany\\
$^6$ Institute of Astronomy, KU Leuven, Celestijnenlaan 200D, 3001 Leuven, Belgium\\
 $^{7}$Royal Observatory of Belgium, Ringlaan 3, B-1180 Brussel, Belgium\\
$^8$Leibniz-Institutfür Astrophysik Potsdam (AIP) An der Sternwarte 16 14482 Potsdam, Germany\\
$^{9}$Dakshin Kamrup College, Gauhati University, Assam-781125-India \\
$^{10}$Department of CSIT, M. J. P. Rohilkhand University, Bareilly, Uttar Pradesh-243006, India\\
$^{11}$Indian Institute of Astrophysics, Koramangala, Bangalore-560034, India\\
 }

\date{Accepted XXX. Received YYY; in original form ZZZ}

\pubyear{2026}

\begin{document}
\label{firstpage}
\pagerange{\pageref{firstpage}--\pageref{lastpage}}
\maketitle

\begin{abstract}
Space-based photometry from \ktwo\ and  \tess, coupled with high-resolution spectroscopy, provides a unique opportunity to re-examine the variability and chemical nature of four bright stars (HD\,73135, BD\,+19\degr\,2045, BD\,+19\degr\,2046, and TYC\,1395-855-1), which were poorly characterised by the original ground-based campaign under the Nainital-Cape survey. We aim to establish the nature of the variability, chemical properties, and evolutionary status of these stars. We analysed ground-based Johnson $BV$ photometry combined with \ktwo\ and \tess\ time-series data to establish the nature of their variability. Using high-resolution HERMES spectroscopy, we determined their spectral classifications and chemical abundances. In addition, we used the inferred spectroscopic constraints with grid-based evolutionary modelling to derive their corresponding masses, radii, and ages. The frequency analysis reveals a diversity of low-amplitude variability across the sample. HD\,73135 shows a persistent modulation near 1.5\,d that is most consistent with rotation, although an ellipsoidal-binary interpretation cannot yet be excluded, while BD\,+19\degr\,2045 exhibits multiple low-frequency signals and is identified as a new candidate $\gamma$\,Doradus $g$-mode pulsator. BD\,+19\degr\,2046 and TYC\,1395-855-1 are non-variable or only marginally variable in \ktwo\ but display coherent low-frequency modulation in \tess\ of uncertain origin. Spectroscopically, HD\,73135 is the only chemically peculiar star in the sample and shows a clear Am abundance pattern, whereas the other three stars are chemically normal. These results demonstrate the value of combining legacy survey data with contemporary photometric and spectroscopic analysis.

\end{abstract}



%


\section{Introduction} \label{sec:intro}

The Nainital--Cape (N--C) Survey was initiated in 1999 as a collaboration between the Aryabhatta Research Institute of Observational Sciences (ARIES), Nainital, India, and the South African Astronomical Observatory (SAAO), South Africa \citep{2000BASI...28..251A}. The survey aimed at searching for and characterising photometric variability, both pulsational and rotational, in candidate chemically peculiar (CP) stars, particularly of the Ap and Am types. Its main results were presented by \citet{2001A&A...371.1048M} and extended by subsequent studies \citep[e.g.][]{2003MNRAS.344..431J, 2016A&A...590A.116J}, with most targets showing no significant pulsational variability.

The N--C survey focused primarily on CP stars, which are upper-main-sequence stars whose spectra display anomalous line strengths and surface abundances relative to normal stars of similar effective temperature \citep{1974ARA&A..12..257P, MichaudAlecianRicher2015}. These peculiarities are understood to arise from processes such as atomic diffusion (including radiative levitation and gravitational settling) in relatively stable stellar atmospheres, and in many cases are linked to large-scale magnetic fields \citep{MichaudAlecianRicher2015}. CP stars are of particular interest because surface abundance inhomogeneities can produce rotational modulation, and some subclasses (especially Am stars) can also exhibit low-amplitude pulsational variability \citep{turcotte2000, 2017MNRAS.465.2662S}.

The N--C sample was assembled by selecting stars with Strömgren $uvby\beta$ indices similar to those of known pulsating Ap and Am stars \citep{2006A&A...455..303J, 2009A&A...507.1763J}. The adopted index ranges 
were chosen to isolate CP late-A to early-F stars, and were slightly broadened (by $\sim$10\%) to include somewhat cooler or more evolved objects \citep{2006A&A...455..303J}. The primary candidate list was drawn from Simbad photometric catalogues \citep{Hauck1998}, and supplemented by catalogues of Ap/Am stars (e.g.\ the Renson and Michigan compilations; see \citealt{2009A&A...507.1763J}). However, the original survey did not include the high-resolution spectroscopy required for full chemical characterisation of the targets, and this effort has therefore been pursued in subsequent phases \citep[e.g.][]{2017MNRAS.467..633J, 2022MNRAS.510.5854J, 2025MNRASJoshi, 2025MNRAS.Dileep}.

The variability of the N--C sample was established based on ground-based Johnson $BV$ time-series photometry. However, ground-based photometric data suffer from diurnal sampling gaps and atmospheric systematics. Therefore, these data are mainly used to place conservative upper limits on the variability and to flag candidate periodicities for verification with space-based light curves.

The advent of long-baseline, high-precision photometry from space missions has transformed the study of stellar variability. The nominal {\it Kepler} mission, for example, revealed coherent low-amplitude signals at previously inaccessible precision \citep{2010Sci...327..977B}. The \ktwo\ mission extended this capability to new sky fields \citep{2014PASP..126..398H, 2022MNRAS.510.5854J}, and \tess\ has continued the legacy with its all-sky surveys \citep{2015JATIS...1a4003R, 2021cosp...43E.499R}. Combined with high-resolution HERMES spectroscopy \citep{2011A&A...526A..69R} and forward-modelling methods \citep{2010aste.book.....A}, such data now allow robust spectral classification, atmospheric parameter determination, and unambiguous interpretation of variability (rotation versus pulsation). This synergy enables a modern re-characterisation of the N--C targets.

The N--C Survey comprises 391 stars, none of which were observed during the nominal \emph{Kepler} mission. However, \ktwo\ data are available for twelve targets. Of these, five were classified as CP (Am) rotational variables by \citet{2022MNRAS.510.5854J}, while three stars (HD\,23734, HD\,68703, and HD\,73345) were identified as chemically normal pulsating variables by \citet{2025MNRASJoshi}. The remaining four stars, HD\,73135, BD\,+19\degr\,2045, BD\,+19\degr\,2046, and TYC\,1395-855-1, form the focus of the present study. For these objects, the availability of \ktwo\ and \tess\ light curves, together with new HERMES spectroscopy, enables a level of characterisation that was not possible at the time of the original survey.

Our observational strategy combines ground-based monitoring with \ktwo\ and \tess\ photometry and new high-resolution HERMES spectroscopy. Using a uniform analysis of all four stars, we aim to (i) search for and characterise low-amplitude variability and assess the consistency of multiple photometric diagnostics, (ii) determine spectral classifications and atmospheric parameters and identify chemical peculiarities where present, and (iii) infer fundamental parameters, including mass, radius, and age, using MESA Isochrones and Stellar Tracks (MIST) evolutionary models \citep{Choi_2016ApJ...823..102C, Dotter_2016ApJS..222....8D}. This study thus provides a consistent modern characterisation of a set of previously poorly studied bright stars in the N--C survey sample, and connects legacy ground-based monitoring with contemporary space-based time-domain photometry.

The paper is structured as follows. Sect.\,\ref{sec:selection} outlines the target selection, observations, and data reduction procedures. Photometric variability analysis is presented in Sect.\,\ref{Nat_Var}, followed by spectral classification in Sect.\,\ref{sec:class}. Sect.\,\ref{para} describes the spectroscopic parameters and abundance determinations, while Sect.\,\ref{evolv} discusses the fundamental parameters and evolutionary interpretation. Our conclusions and prospects are summarised in Sect.\,\ref{conclusion}.

\section{Target sample and observations}
\label{sec:selection}

The four objects (HD\,73135, BD\,+19\degr\,2045, BD\,+19\degr\,2046, and TYC\,1395-855-1) define a quartet that was selected from the N--C survey sample to meet the following criteria: (i) availability of long- and/or short-cadence \ktwo\ and \tess\ photometry (Table\,\ref{tab:campaign}), (ii) comparable brightness ($V\simeq8.6$–10.4\,mag) to ensure uniform photometric precision, (iii) low line-of-sight extinction ($A_V<0.15$\,mag), and (iv) closely matched ecliptic latitudes to minimise differential roll systematics and facilitate homogeneous reduction. The brightness of the targets ($V\simeq 8.6$--10.4\,mag) makes them readily accessible to 1--2\,m class telescopes.   

The full data set therefore consists of (i) ground-based CCD photometry obtained with 0.4--1.3\,m class telescopes, (ii) long-baseline \ktwo\ and \tess\ space photometry, and (iii) high-resolution optical spectroscopy. The basic instrumental characteristics are summarised in Table\,\ref{table:tele}, and a detailed observing log of the spectroscopic material is given in Table\,\ref{table:log}. Importantly, each star is observed in multiple independent data sets (ground-based, \ktwo, and \tess) with different systematics. Comparing these data sets provides a robust check on signal persistence and supports a more conclusive interpretation of the variability.

\begin{table}
\caption{Overview of \ktwo\ campaigns (C05 and C18) and \tess\ sectors (S45 and S72) in which the programme stars were observed with the corresponding cadence.}
\label{tab:campaign}
\centering
\begin{tabular}{lcccc}
\hline
\noalign{\smallskip}
Mission         & \ktwo     & \ktwo     & \tess     & \tess     \\
Campaign/Sector & C05       & C18       & S45       & S72       \\
Start date      & 27/04/15  & 12/05/18  & 06/11/21  & 11/11/23  \\
End date        & 10/07/15  & 02/07/18  & 02/12/21  & 07/12/23  \\
\noalign{\smallskip}\hline\noalign{\smallskip}
HD\,73135       &     1800 s &    1800 s       &      475 s &      158 s \\
\noalign{\smallskip}
BD\,+19\degr\,2045 &  1800 s &     1800 s &           &      158 s \\
                &       60 s &           &           &           \\
\noalign{\smallskip}
BD\,+19\degr\,2046 &  1800 s &           &      475 s &      158 s \\
\noalign{\smallskip}
TYC\,1395-855-1 &     1800 s &     1800 s &           &      158 s \\
\noalign{\smallskip}\hline
\end{tabular}
\end{table}


\begin{table*}
\caption{Specifications of the telescopes and backend instruments used for the photometric and spectroscopic observations.}
\label{table:tele}
\centering
\fontsize{7.2}{9.0}\selectfont
\begin{tabular}{lllll l}
\hline
\noalign{\smallskip}
Category & Facility & Diameter (cm) & Location & Detector/Spectrograph & FoV / Resolution \\
\noalign{\smallskip}\hline\noalign{\smallskip}
Ground-based photometry
& DFOT            & 130 & Devasthal (India)                 & Andor DZ436                 & $18^{\prime}\!\times18^{\prime}$ \\
&                &     &                                   & Andor iXon EM+ DU-897        & $4.7^{\prime}\!\times4.7^{\prime}$ \\
& ST              & 104 & Nainital (India)                  & 2k Wright CCD               & $13^{\prime}\!\times13^{\prime}$ \\
& MASTER-II-URAL  & 40  & Ural Federal University (Russia)  & U16M CCD                     & $2^{\circ}\!\times4^{\circ}$ \\
& PROMPT-8        & 60  & CTIO (Chile)                      & Apogee F42                   & $26.6^{\prime}\!\times22.6^{\prime}$ \\
\noalign{\smallskip}\hline
Space-based photometry
& \ktwo           & 95  & Space                             & ---                          & $10.25^{\circ}\!\times10.25^{\circ}$ \\
& \tess           & 10.5 (per camera; 4 cameras) & Space                             & ---                          & $24^{\circ}\!\times96^{\circ}$ \\
\noalign{\smallskip}\hline
Spectroscopy
& Mercator        & 120 & La Palma (Canary Islands, Spain)                         & HERMES                       & ${\cal R} \simeq 85\,000$ \\
\noalign{\smallskip}\hline
\end{tabular}
\end{table*}

\begin{table*}
\caption{The HERMES observation log for the target stars.  The right ascension ($\alpha_{\rm J2000}$) and declination ($\delta_{\rm J2000}$) given by \citet{Gaia3} and the V-band magnitude estimated by \citet{2000A&A...355L..27H} are included.}
\label{table:log}
\begin{tabular}{llcccccccr}
\hline\hline\\
TIC&Other&$\alpha_{\rm J2000}$&$\delta_{\rm J2000}$ & V & BJD (2\,460\,000+)&Number& SNR&  Total Exposure\\
No.&Name&(hh mm ss)&(dd mm ss)&(mag)&(day)&of&(@ 500, 650, &Duration\\
&&&&&&Spectra& and 810 nm)&(min)\\
\noalign{\smallskip}
\hline
\noalign{\smallskip}
  195192995  & HD\,73135 & 08 37 18.2273&+18 52 11.8914 & 8.59  & 780  &1& 108, 113, 82 & 15\\
  195192330  & BD\,+19\degr\,2045 & 08 36 29.8499& +18 57 57.0725& 9.50  & 780   & 1& 121, 113, 82 & 35\\
  195192363  & BD\,+19\degr\,2046 & 08 36 34.5373&+18 51 13.2893 & 9.82 & 780  & 1& 120, 59, 50 & 45\\
  195193042 & TYC\,1395-855-1 & 08 37 01.6850&+19 01 38.6082& 10.38 & 781 &2& 93, 73, 47 & 90\\
\hline
\end{tabular}
\end{table*}

\subsection{Ground‑based observations}
\label{sec:ground}

Johnson $B$ and $V$ time‑series of HD\,73135 and BD\,+19\degr\,2045 were obtained with the 1.04-m Sampurnanand Telescope at Nainital Observatory (ARIES; India), the 1.3-m Devasthal Fast Optical Telescope at Devasthal Observatory (ARIES; India), the 0.60-m Thai Southern Hemisphere Telescope (PROMPT-8) at Cerro Tololo Observatory (NARIT; Chile), and the 0.4-m MASTER-II-Ural telescope at Kourovka Observatory (Ural Federal University; Russia) \citep{2010AdAst2010E..30L, 2014AstBu..69..368B}.  Standard bias, dark, and twilight‑flat frames were obtained on a nightly basis. The preliminary data reduction was performed using \textsc{iraf}\footnote{\textsc{iraf} is distributed by the National Optical Astronomy Observatory, which is operated by the Association of Universities for Research in Astronomy under a cooperative agreement with the National Science Foundation.} and it included bias/dark subtraction, flat‑fielding, cosmic‑ray cleaning, and aperture photometry with \textsc{daophot} \citep{Stetson1987}. BD\,+19\degr\,2046 and TYC\,1395-855-1 were used as comparison stars for the differential photometry. 

\begin{figure}
\centering
 \includegraphics[width=\columnwidth]{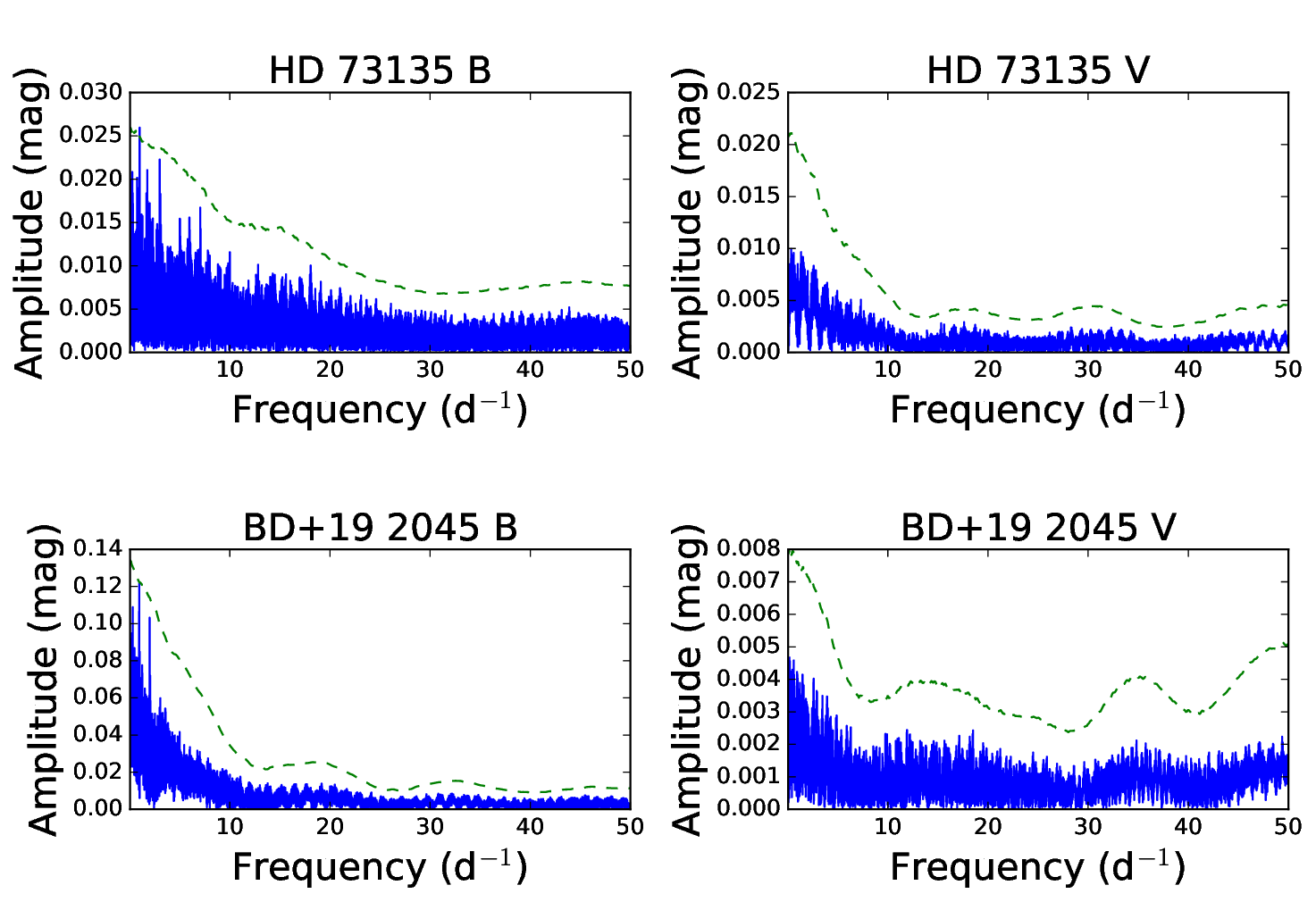}
\caption{Lomb–Scargle periodograms of the ground‑based Johnson $B$ (left) and $V$ (right) light curves. Top panels: HD\,73135; bottom  panels: BD\,+19\degr\,2045. The ordinate is amplitude (magnitudes); the dashed line marks the ${\rm SNR}=4.0$ detection threshold, computed from a running-median noise estimate. Only the \textit{B}-band spectra show peaks that reach the threshold, clustered near $\sim$1–2\,d$^{-1}$ (and harmonics) consistent with diurnal sampling/colour-dependent extinction; no significant periodicities are detected in \textit{V}.}
   \label{ground_ob}
\end{figure}

The light curves were analysed with Lomb--Scargle (LS) periodograms \citep{1976Ap&SS..39..447L, 1982ApJ...263..835S}. The average amplitude of the residual spectrum after pre-whitening provides a proxy for the local noise.  Following the commonly used criterion of \citet{1993A&A...271..482B}, we adopt ${\rm SNR}\ge4.0$ as the significance threshold for independent frequencies; peaks with $3.5\le{\rm SNR}<4.0$ are treated as tentative \citep{1999A&A...349..225B}. Fig.\,\ref{ground_ob} displays the resulting periodograms, where the dashed line marks the threshold ${\rm SNR}=4.0$. 

Considering the tentative level ($3.5\le{\rm SNR}<4.0$), peaks appear only in the \textit{B}-band periodograms, with none in \textit{V} reaching ${\rm SNR}\ge3.5$. For HD\,73135 (\textit{B}), we find frequency peaks  at $f\simeq 0.994$, $1.756$, $2.997$, and $7.046~\mathrm{d}^{-1}$ (periods $\approx 1.006$, $0.570$, $0.334$, and $0.142~\mathrm{d}$) with amplitudes $\approx 0.017$–$0.026~\mathrm{mag}$, while BD\,+19\degr\,2045 (\textit{B}) shows frequencies $f\simeq 0.954$ and $1.993\,{\rm d}^{-1}$ (periods $\approx 1.048$ and $0.502~\mathrm{d}$) with larger amplitudes $\approx 0.10$–$0.12\,\mathrm{mag}$. Peaks near $\sim 1$ and $\sim 2~\mathrm{d}^{-1}$ (and related harmonics, e.g. $\sim 3~\mathrm{d}^{-1}$) are emblematic of diurnal window-function effects and differential extinction in ground-based data; their occurrence exclusively in \textit{B} and absence in \textit{V} therefore point to a non-astrophysical origin. For HD\,73135, the peak at $7.046\,\mathrm{d}^{-1}$ lies in a $\delta$\,Scuti-like frequency range but is borderline in significance and uncorroborated in \textit{V}, and should be regarded as tentative.  Overall, none of the peaks constitutes a secure stellar periodicity; a direct comparison with the space-based \ktwo\ and \tess\ periodograms is presented in Sect.\,\ref{Nat_Var} (see also Tables\,\ref{table:KepWhite2015}--\ref{tab:tess}).

\subsection{\ktwo\ space photometry}%
\label{sec:obs:k2}

 The \ktwo\ data comprise Campaigns\,5 and 18 for HD\,73135, BD\,+19\degr\,2045, and TYC\,1395-855-1, while BD\,+19\degr\,2046 is covered only in Campaign\,5 
(Table\,\ref{tab:campaign}). In this paper, we describe the analysis of the long cadence data ($\sim$ 29.4-min) for all stars; however, a similar analysis was employed later for the short-cadence data ($\sim$ 58.8\,s) available for BD+19\degr\,2045.  Long‑cadence light curves were downloaded from MAST as Pre‑search Data Conditioning Simple Aperture Photometry (PDCSAP) fluxes\footnote{\url{https://archive.stsci.edu/k2/}}. For each campaign, we also obtained the K2 Systematics Correction (K2SC) products \citep{2015MNRAS.447.2880A}, which remove the dominant roll‑angle drift trend.
 
Visual inspection showed that PDCSAP and K2SC curves agree within $\lesssim$300\,ppm for all stars. Since K2SC occasionally introduces low‑order trends in very quiet targets, we adopted PDCSAP fluxes as the default data set and used K2SC only as a consistency check. Fluxes were converted to relative units, $\Delta F_{i} = F_{i}/\langle F\rangle - 1$, and the time series were shifted in such a way that $t(i) = t - t_{0}$, where $t_{0}$ was chosen as the time of the first observation in the corresponding light curve ($t_{0}=245\,7139.63$\,d for C5 and $245\,8251.55$\,d for C18).

\begin{figure*}
\includegraphics[width=\textwidth]{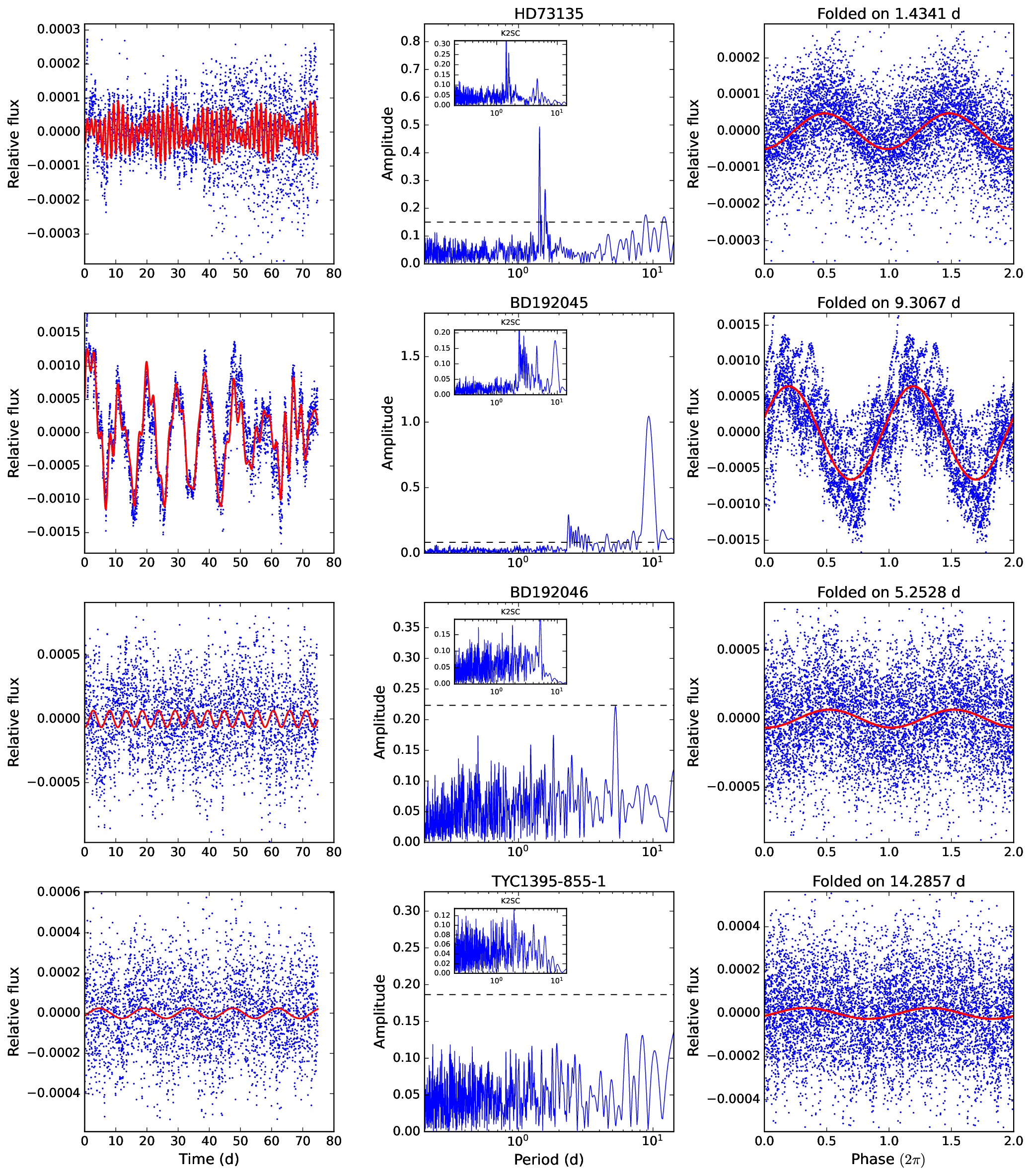}
\caption{{\it Left panels:} Long cadence PDC light curves ({\it blue dots}) are overplotted with models including the significant frequencies (red line) for the four stars studied in this paper during the 2015 K2 Campaign 5. {\it Middle panel:} Lomb-Scargle periodograms. The period distribution ({\it blue}) has been plotted on a log-linear plot to aid visualization. A signal-to-noise ratio (SNR) of 4.0 is indicated by the black dashed horizontal line. The inset shows the Lomb-Scargle periodogram relating to the time-series after the application of the K2SC algorithm. {\it Right panel:} Campaign 5 time-series data ($blue$) phase-folded to the dominant period. The model for the dominant component ($red$) is produced from the data in Table\,\ref{table:KepWhite2015}. }
\label{K2_2015_Periodograms}
\end{figure*}

The LS periodograms (Fig.\,\ref{K2_2015_Periodograms}) were computed up to the Nyquist frequency limit of 24.3\,d$^{-1}$. A frequency peak was considered significant only if it appeared consistently in both observing campaigns (when available), in both the PDCSAP and K2SC light-curve extractions, and satisfied the false-alarm probability criterion ${\rm FAP}\leq10^{-8}$ \citep{2022MNRAS.510.5854J}.

For each extracted frequency $f_k$, we define the phase as:
\begin{equation}
\phi_k(t)=\mathrm{mod}\!\left[(t-t_0)\,f_k,\,1\right].
\end{equation}
The phases reported in Tables\,\ref{table:KepWhite2015} and \ref{table:KepWhite2018} correspond to the fitted phase offsets of the sinusoidal model at the reference epoch $t_0$. Phase-folded light curves are computed using the same definition, adopting $f_{\rm dom}$ for the dominant periodicity shown in each panel.

The detected frequencies are listed in Tables\,\ref{table:KepWhite2015} and\,\ref{table:KepWhite2018}. For all peaks above the adopted detection threshold, we applied iterative sinusoidal fitting and subtraction (pre-whitening), estimating formal uncertainties following \citet{1999DSSN...13...28M}. Figs.\,\ref{K2_2015_Periodograms}, \ref{K2_2018_Periodograms}, and \ref{BD+192045_sc} display the K2 light curves overlaid with the multi-frequency model fits, together with the corresponding LS periodograms and phase-folded light curves for the dominant periodicities. For visualisation, phase-folded light curves are shown over two consecutive cycles ($0 \le \phi < 2$).

\begin{table*}
\centering
\caption{Sinusoidal components identified in the \ktwo\ C5 light curves observed in 2015. Frequencies ($f$), periods ($P$), fractional-deviation semi-amplitudes ($A$), and phases ($\phi$) were obtained by iterative pre-whitening (Sect.\,\ref{sec:obs:k2}). Phases are given relative to the reference epoch $t_{0}$, chosen as the time of the first observation in each time-series data set. The final column gives $\log_{10}({\rm FAP})$, the base-10 logarithm of the false-alarm probability; more negative values denote more significant detections (our acceptance criterion is $\log_{10}{\rm FAP}\le-8$). One unit in $A$ corresponds to $1086$\,mmag, and for each star the components are listed in order of decreasing $A$. }
\label{table:KepWhite2015}
\begin{tabular}{lcccccc}
\hline
\noalign{\smallskip}
Star & $f$ (d$^{-1}$) & $P$ (d) & $A$ $(\times10^{-5})$ & $\phi$ (rad) & $\log_{10}\mathrm{FAP}$ \\
\noalign{\smallskip}\hline\noalign{\smallskip}
HD\,73135 &&&&&\\
             & $0.6973 \pm 0.0005$ & $1.434 \pm 0.001$ & $4.6 \pm 0.3$ & $-0.90 \pm 0.07$ & $-69$\\
             & $0.6333 \pm 0.0009$ & $1.579 \pm 0.003$ & $2.4 \pm 0.3$ & $1.20 \pm 0.18$  & $-18$\\[2pt]
BD\,+19\degr\,2045&&&&&\\
             & $0.1074 \pm 0.0002$ & $9.307 \pm 0.020$ & $64 \pm 2$   & $0.47 \pm 0.02$ & $-1000$\\
             & $0.4257 \pm 0.0004$ & $2.349 \pm 0.002$ & $17.7 \pm 0.9$ & $-1.46 \pm 0.05$ & $-28$\\
             & $0.1192 \pm 0.0004$ & $8.392 \pm 0.028$ & $13.8 \pm 0.8$ & $1.66 \pm 0.06$  & $-11$\\
             & $0.0947 \pm 0.0005$ & $10.56 \pm 0.05$  & $12.1 \pm 0.7$ & $0.11 \pm 0.06$  & $-10$\\[2pt]
BD\,+19\degr\,2046&&&&&\\
             & $0.0523 \pm 0.0008$ & $19.12 \pm 0.29$ & $6.8 \pm 0.7$ & $2.9 \pm 0.1$ & $-16$\\
             & $0.1900 \pm 0.0008$ & $5.253 \pm 0.024$ & $6.7 \pm 0.7$ & $-0.9 \pm 0.1$& $-14$\\[2pt]
TYC\,1395‑855‑1&-&-&-&-&-\\
\noalign{\smallskip}\hline
\end{tabular}
\end{table*}


\begin{table*}
\centering
\caption{Same as Table\,\ref{table:KepWhite2015} but for \ktwo\ Campaign 18 (2018).}
\label{table:KepWhite2018}
\begin{tabular}{lcccccc}
\hline
\noalign{\smallskip}
Star & $f$ (d$^{-1}$) & $P$ (d) & $A$ $(\times10^{-5})$ & $\phi$ (rad) & $\log_{10}\mathrm{FAP}$ \\
\noalign{\smallskip}\hline\noalign{\smallskip}
HD\,73135 &&&&&\\
             & $0.651 \pm 0.001$ & $1.537 \pm 0.003$ & $2.9 \pm 0.3$ & $-3.10 \pm 0.10$ & $-22$\\
             & $0.056 \pm 0.001$ & $17.8 \pm 0.4$    & $2.8 \pm 0.3$ & $-1.80 \pm 0.10$ & $-20$\\
             & $3.860 \pm 0.002$ & $0.2591 \pm 0.0001$ & $1.7 \pm 0.3$ & $-2.7 \pm 0.2$  & $-11$\\
             & $0.427 \pm 0.002$ & $2.34 \pm 0.01$   & $1.6 \pm 0.3$ & $-1.2 \pm 0.2$   & $-10$\\
             & $0.499 \pm 0.002$ & $2.003 \pm 0.007$ & $1.6 \pm 0.3$ & $-1.3 \pm 0.2$   & $-9$\\[2pt]
BD\,+19\degr\,2045&&&&&\\
             & $0.1151 \pm 0.0004$ & $8.687 \pm 0.028$ & $50 \pm 2$   & $1.53 \pm 0.03$ & $-221$\\
             & $0.2215 \pm 0.0004$ & $4.515 \pm 0.009$ & $30 \pm 2$   & $-1.80 \pm 0.04$ & $-95$\\
             & $0.3031 \pm 0.0006$ & $3.299 \pm 0.006$ & $17.3 \pm 0.9$ & $-0.96 \pm 0.05$ & $-34$\\
             & $0.0796 \pm 0.0007$ & $12.56 \pm 0.11$ & $16 \pm 1$   & $2.29 \pm 0.06$ & $-22$\\
             & $0.4050 \pm 0.0006$ & $2.469 \pm 0.004$ & $13.8 \pm 0.8$ & $1.69 \pm 0.06$ & $-21$\\
             & $0.1009 \pm 0.0008$ & $9.91 \pm 0.08$  & $12.7 \pm 0.9$ & $0.04 \pm 0.08$ & $-16$\\
             & $0.1264 \pm 0.0008$ & $7.91 \pm 0.05$  & $9.8 \pm 0.7$ & $-2.60 \pm 0.08$ & $-13$\\
             & $0.3273 \pm 0.0009$ & $3.055 \pm 0.008$ & $8.2 \pm 0.7$ & $1.94 \pm 0.08$ & $-12$\\[2pt]
TYC\,1395‑855‑1&&&&&\\
             & $0.100 \pm 0.002$  & $10.0 \pm 0.2$   & $4.1 \pm 0.5$ & $2.7 \pm 0.2$ & $-9$\\
\noalign{\smallskip}\hline
\end{tabular}
\end{table*}

\subsection{\tess\ Photometry}
 The \tess\ data span Sectors\,44, 45, 46, and 72; however, Sectors\,44 and 46 are strongly affected by instrumental artefacts for all targets, and Sector\,45 is likewise affected for BD\,+19\degr\,2045 and TYC\,1395-855-1. We therefore exclude these affected sectors from our analysis. The combined baselines and cadences provide sensitivity to rotational modulation and low-frequency pulsations, if present. 

We extracted aperture photometry from the \tess\ Full-Frame Images (FFIs) for the usable sectors listed in Sect.\,\ref{sec:selection} and Table\,\ref{tab:campaign}. The light curves were produced using a consistent extraction and cleaning procedure, as described below.

An optimal elliptical photometric aperture was determined for each target by minimising the median absolute deviation (MAD) of the resulting light curve, thereby maximising the inclusion of target flux relative to background noise. Robust background subtraction was applied using a local annular background mask to estimate and remove the sky contribution. Outlier flux measurements were identified and removed through iterative sigma clipping.

All data processing was performed using the \texttt{lightkurve} package \citep{2018ascl.soft12013L} in conjunction with the \texttt{astropy} library \citep{astropy:2013, astropy:2018, astropy:2022}. Periodograms (see Fig.\,\ref{fig:ft_tess}) were computed and significant periodic signals were identified following the methodology described in Sect.\,\ref{sec:obs:k2}. The properties of the detected significant frequencies are summarised in Table\,\ref{tab:tess}.

\begin{table*}
\caption{ Same as Table\,\ref{table:KepWhite2015}, but for \tess\ data. Frequencies ($f$), periods ($P$), fractional-deviation semi-amplitudes ($A$), phases ($\phi$), and false-alarm probabilities are listed for the significant signals detected in the \tess\ light curves. Phases are given relative to the reference epoch $t_{0}$, chosen as the time of the first observation in each sector (see Sect.\,\ref{sec:obs:k2}).}
\label{tab:tess}
\begin{tabular}{lccccccc}
\hline
\noalign{\smallskip}
Star &Sector & $f$ (d$^{-1}$) & $P$ (d) & $A$ $(\times10^{-5})$ & $\phi$ (rad) & $\log_{10}\mathrm{FAP}$ \\
\noalign{\smallskip}\hline\noalign{\smallskip}
  HD\,73135 & 45 & 0.229 $\pm$ 0.002 & 4.37 $\pm$ 0.04 & 8.5 $\pm$ 0.70 & 2.457 $\pm$ 0.014 & -17\\
  \noalign{\smallskip}
   & 72 & 0.235 $\pm$ 0.002 & 4.26 $\pm$ 0.03 & 8.7 $\pm$ 0.60 & 1.366 $\pm$ 0.011 & -36\\
   && 0.311 $\pm$ 0.002 & 3.21 $\pm$ 0.02 & 6.3 $\pm$ 0.60 & 0.485 $\pm$ 0.015 & -14\\
   \hline
  BD+19\degr\,2045 & 72 & 0.102 $\pm$ 0.001 & 9.83 $\pm$ 0.10 & 17.5 $\pm$ 0.80 & 1.829 $\pm$ 0.007 & -101\\
   && 0.229 $\pm$ 0.001 & 4.38 $\pm$ 0.02 & 16.8 $\pm$ 0.70 & 2.2 $\pm$ 0.007 & -92\\
   && 0.392 $\pm$ 0.001 & 2.55 $\pm$ 0.01 & 14.7 $\pm$ 0.70 & 2.352 $\pm$ 0.008 &-68\\
   && 0.273 $\pm$ 0.002 & 3.66 $\pm$ 0.03 & 8.8 $\pm$ 0.70 & 0.064 $\pm$ 0.013 & -18\\
   && 0.196 $\pm$ 0.002 & 5.10 $\pm$ 0.06 & 7.3 $\pm$ 0.70 & 0.812 $\pm$ 0.016 & -9\\
\hline
  BD+19\degr\,2046& 45 & 0.166 $\pm$ 0.002 & 6.04 $\pm$ 0.08 & 11.5 $\pm$ 1.1 & 0.462 $\pm$ 0.016 & -11\\
\noalign{\smallskip}
   &72 & 0.155 $\pm$ 0.002 & 6.40 $\pm$ 0.06 & 14.9 $\pm$ 1.0 & 1.057 $\pm$ 0.011 & -33\\
   \hline
  TYC\,1395-855-1 &72 & 0.155 $\pm$ 0.001 & 6.47 $\pm$ 0.04 & 30.9 $\pm$ 1.2 & 0.61 $\pm$ 0.006 & -124\\
   && 0.229 $\pm$ 0.001 & 4.37 $\pm$ 0.02 & 21.7 $\pm$ 1.2 & 1.075 $\pm$ 0.009 & -56\\
   && 0.462 $\pm$ 0.002 & 2.17 $\pm$ 0.01 & 12.2 $\pm$ 1.2 & 1.821 $\pm$ 0.015 & -11\\
   && 0.300 $\pm$ 0.002 & 3.34 $\pm$ 0.02 & 12.3 $\pm$ 1.2 & 1.509 $\pm$ 0.015 & -11\\
  \hline\end{tabular}
\end{table*}

\subsubsection{Time–frequency Analysis Toolkit}
\begin{figure*}
\centering
 \includegraphics[width=\textwidth]{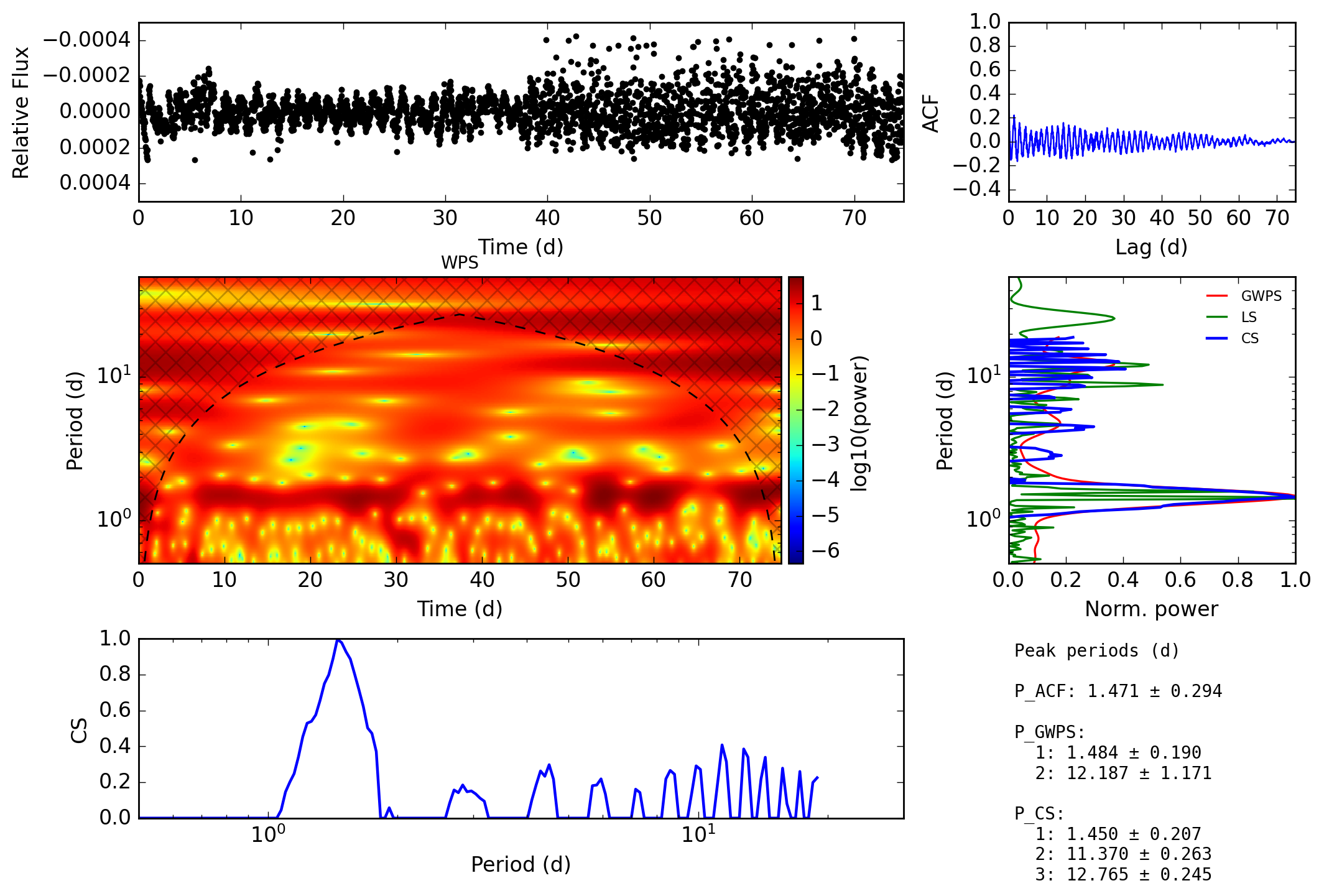} 
\caption{ Period diagnostics for the \ktwo\ C05 light curve of HD\,73135. Top: light curve flux (left) and autocorrelation function (ACF; right). Middle-left: Morlet wavelet power spectrum (log$_{10}$ power) computed on 200 logarithmically spaced periods spanning 0.5--30\,d using symmetric reflection padding; the dashed curve marks the cone of influence (COI; \citealt{TorrenceCompo1998}) and the hatched region indicates where edge effects become important. Middle-right: global wavelet power spectrum (GWPS; red), Lomb--Scargle periodogram evaluated on the same period grid (LS; green), and their composite spectrum (CS; blue), defined as the geometric mean of the normalised GWPS and ACF \citep{2016MNRAS.456..119C, 2017A&A...605A.111C}. Bottom-left: CS versus period. Bottom-right: dominant period from the first ACF peak ($P_{\rm ACF}$) and the strongest peaks in GWPS and CS ($P_{\rm GWPS}$, $P_{\rm CS}$); uncertainties are estimated from peak half-height widths ($\sigma_P \simeq {\rm FWHM}/2.355$).}
\label{wavel2} 
\end{figure*}

All wavelet power spectra were computed  using a custom python implementation of the Morlet wavelet transform \citep{TorrenceCompo1998}, on a logarithmic grid of 200 scales spanning 0.5–80\,d.  We applied symmetric reflection padding to minimise edge effects and shade the cone of influence as defined in \citet{TorrenceCompo1998}.  Global wavelet power spectra (GWPS) were obtained by summing the wavelet power over time at each scale and normalised to unit variance following \citet{Liu2007}.  Autocorrelation functions were computed from linearly detrended \ktwo\ and \tess\  light curves. The composite spectrum (CS) is defined as the geometric mean of the normalised GWPS and the  autocorrelation function (ACF), following \citet{2016MNRAS.456..119C, 2017A&A...605A.111C}, thereby emphasising periodicities that are both strong and temporally persistent. The results are presented in Figs.\,\ref{wavel2} and \ref{wavel3}–\ref{wavel15}.

\subsection{High‑resolution spectroscopy}%
\label{sec:spec}

High‑resolution spectra of all four stars were obtained with the HERMES \'{e}chelle spectrograph \citep{2011A&A...526A..69R} equipped on the 1.2‑m Mercator Telescope (La Palma) on Apr\,14–15, 2025. In a single exposure, HERMES covers a wavelength range of 377–900\,nm  with spectral resolution of ${\cal R}\approx85\,000$ and stability $\sim$50\,m\,s$^{-1}$.  The dedicated pipeline includes bias subtraction, flat‑fielding, optimal order extraction, wavelength calibration using Th–Ar lamps, and blaze correction.  Orders were merged and spectra were continuum normalised via low‑order spline fits guided by templates from the Melchior library \citep{2024A&A...681A.107R}.  Barycentric corrections were applied to each spectrum.  Two exposures of TYC\,1395‑855‑1 were median‑combined, yielding ${\rm SNR}\simeq85$ at 550\,nm; single‑exposure ${\rm SNR}$ values for the other stars range from 47 to 121. The resulting spectra form the basis for the spectral classification (see Sect.\,\ref{sec:class}), atmospheric parameter, and abundance analyses (see Sect.\,\ref{para}).

\section{Nature of Variability}
\label{Nat_Var}

Our four programme stars exhibit three qualitatively different kinds of variability in the space-based photometry. Because their precise evolutionary status is established only after the spectroscopic analysis and forward modelling described in Sects.\,\ref{para} and \ref{evolv}, we include a forward reference in parentheses at the first mention of each classification.

A direct comparison with the space photometry clarifies the status of the candidate peaks found in the ground-based data (Sect.\,\ref{sec:ground}). The dominant periodicities identified in the \ktwo\ and \tess\ light curves for all target stars occur at much lower frequencies than the ground-based candidates: HD\,73135 shows a stable signal at $f\simeq0.63$--0.70\,d$^{-1}$ ($P\simeq1.4$--1.6\,d) in \ktwo\ Campaigns\,5 and 18 (Tables\,\ref{table:KepWhite2015} and \ref{table:KepWhite2018}), while BD\,+19\degr\,2045 is dominated by $f\simeq0.10$--0.12\,d$^{-1}$ ($P\simeq9$--10\,d) in both \ktwo\ and \tess\ (Tables\,\ref{table:KepWhite2015}, \ref{table:KepWhite2018}, and \ref{tab:tess}). None of the prominent ground-based peaks near $\sim$1 and $\sim$2\,d$^{-1}$ is recovered as a dominant, persistent signal in the space-based periodograms, consistent with their interpretation as sampling/extinction artefacts. 

A frequency is regarded as significant only if the LS peak satisfies $\log_{10}({\rm FAP})\le -8$ (approximately a $\sim4\sigma$ detection) following \citet{2022MNRAS.510.5854J}. Because space-based periodograms evaluate a very large number of trial frequencies and may retain low-level instrumental systematics, we adopt this conservative threshold to define significant peaks. This criterion is intentionally more stringent than the ground-based amplitude SNR thresholds (Sect.\,\ref{sec:ground}), which are tailored to shorter, window-function--limited time series and are used there mainly to flag candidate peaks. Moreover, no single diagnostic is sufficient for low-frequency variability: the LS periodogram alone can be sensitive to residual systematics and harmonics, while real signals due to rotation or pulsation may vary in amplitude and phase with time. We therefore require \emph{consistency} among four complementary diagnostics:
\begin{enumerate}
\item the LS/pre-whitened frequency list (Tables\,\ref{table:KepWhite2015} and \ref{table:KepWhite2018});
\item the wavelet power spectrum (WPS $\rightarrow$ GWPS), which distinguishes spot-like rotational modulation (a single drifting ridge) from pulsational behaviour (multiple quasi-stationary ridges);
\item the autocorrelation function (ACF), whose first secondary maximum yields an independent rotational period, $P_{\rm ACF}$;
\item the composite spectrum (CS), defined as the geometric mean of GWPS and ACF \citep{2016MNRAS.456..119C, 2017A&A...605A.111C}, which enhances periodicities that are both strong and time-persistent.
\end{enumerate}

We classify the variability as {\it rotational} if (a) $P_{\rm LS}$, $P_{\rm ACF}$, $P_{\rm GWPS}$, and $P_{\rm CS}$ agree within their uncertainties; (b) the photometric period is compatible with the spectroscopic constraint $P_{\rm phot}\le P_{\max}\equiv 2\pi R/(v\sin i)$, where $R$ is taken from Sect.\,\ref{evolv}; and (c) the WPS ridge broadens with time in a manner typical of spot evolution and/or differential rotation. We deem the variability of {\it pulsational} origin if multiple harmonically unrelated peaks cluster in the $\gamma$\,Doradus domain ($0.3$--$3$\,d$^{-1}$), form narrow, quasi-stationary ridges in the WPS, and do not produce a coherent ACF signature.

\begin{table*}
\centering
\caption{Summary of variability diagnostics. Columns (3--7): dominant significant period from the composite spectrum ($P_{\rm CS}$), semi-amplitude of the strongest LS peak ($A_{\max}$; 1 unit in $A$ corresponds to 1086\,mmag), range of $\log_{10}({\rm FAP})$, a concordance flag, and the variability classification. Concordance flag: ``Y'' = all four metrics agree; ``N'' = at least one metric disagrees or $\log_{10}{\rm FAP}>-8$.}
\label{tab:var_summary}
\setlength{\tabcolsep}{3pt}
\renewcommand{\arraystretch}{0.95}
\begin{tabular}{lcccccc}
\hline
Star & K2 Campaign(s)/Sector(s) & Dominant $P_{\rm CS}$ (d) & $A_{\max}$ (mmag) & $\log_{10}{\rm FAP}$ range & Concordance & Classification \\
\hline
HD\,73135          & 5, 18   & 1.45$\pm$0.21; 1.55$\pm$0.26 & 0.05   & $-69$ to $-9$    & Y & Rotation (surface inhomogeneities) \\
BD\,+19\degr\,2045 & 5, 18   & 9.45$\pm$1.18; 8.82$\pm$1.07   & 0.70   & $-1000$ to $-10$ & Y & $\gamma$\,Doradus $g$-modes \\
BD\,+19\degr\,2046 & 5       & 5.06$\pm$0.17                & $<0.07$ & $-16$ to $-5$   & N & Non-significant \\
TYC\,1395$-$855$-$1 & 5, 18  & 18.06$\pm$0.33; 10.1$\pm$0.50 & $<0.04$ & $-9$ to $-4$    & N & Non-significant \\
\hline
HD\,73135          & 45, 72  & 4.51$\pm$0.39; 3.49$\pm$0.55   & 0.09   & $-36$ to $-14$  & N & Long-period signal (inconsistent with $v\sin i$) \\
BD\,+19\degr\,2045 & 72      & 9.9$\pm$0.68                 & 0.20   & $-101$ to $-9$  & Y & $\gamma$\,Doradus $g$-modes \\
BD\,+19\degr\,2046 & 45, 72  & 4.11$\pm$0.05; 6.38$\pm$0.29   & $<0.16$ & $-33$ to $-11$ & Y & Low-frequency variability (uncertain) \\
TYC\,1395$-$855$-$1 & 72     & 7.33$\pm$1.22                 & $<0.34$ & $-124$ to $-11$ & Y & Low-frequency variability (uncertain) \\
\hline
\end{tabular}
\end{table*}

\subsection{Star-by-star classification from \ktwo\ and \tess\ Photometry}
\label{Nat_Var_starby}

In \ktwo\ Campaigns\,5 and 18, HD\,73135 shows a persistent low-frequency doublet at $f\simeq0.63$--$0.70$\,d$^{-1}$ ($P\simeq1.4$--1.6\,d) with highly significant detections (down to $\log_{10}{\rm FAP}\simeq-69$). The agreement between the LS, ACF, GWPS, and CS diagnostics, together with the broadening of the WPS ridge, supports rotational modulation by evolving surface inhomogeneities (e.g.\ abundance spots) and/or differential rotation. Using the radius from Sect.\,\ref{evolv} and the measured $v\sin i$ (Table\,\ref{tab:sspec_para}), we obtain $P_{\max}=2\pi R/(v\sin i)\approx1.8$\,d, confirming that the $\sim$1.5\,d modulation is compatible with rotation for a moderate inclination. 
In \tess\ (Sectors\,45 and 72) the dominant low-frequency power shifts to longer timescales ($P\simeq3.2$--4.4\,d; Table\,\ref{tab:tess}), which exceed $P_{\max}$ and therefore cannot represent the fundamental rotational period. We adopt the K2 $\sim$1.5\,d signal as the rotational timescale. Photometry alone cannot strictly exclude an orbital interpretation (ellipsoidal variability in a non-eclipsing binary), for which the dominant photometric signal typically occurs at $2f_{\rm orb}$; a dedicated radial-velocity time series is required to rule out this possibility.

BD\,+19\degr\,2045 shows the most complex low-frequency variability in both \ktwo\ and \tess. In \ktwo\ C5, it is dominated by $f\simeq0.108$\,d$^{-1}$ ($P\simeq9.30$\,d), with additional significant peaks spanning $\sim0.09$--$0.43$\,d$^{-1}$; \ktwo\ C18 and \tess\ Sector\,72 show a rich but comparable pattern across $\sim0.08$--$0.41$\,d$^{-1}$, producing a quasi-stationary multi-ridge WPS morphology. Although the strongest periods are of order 9--10\,d, the presence of multiple harmonically unrelated peaks and persistent, narrow ridges is consistent with $\gamma$\,Doradus $g$-modes rather than spot modulation. Crucially, a rotational origin is excluded by the spectroscopic constraint: the radius from Sect.\,\ref{evolv} and $v\sin i$ (Table\,\ref{tab:sspec_para}) imply $P_{\max}\approx5.7$\,d, which is incompatible with the observed $\sim$9--10\,d periodicity. We therefore classify BD\,+19\degr\,2045 as a new candidate $\gamma$\,Doradus $g$-mode pulsator.

In \ktwo\ C5 data, BD\,+19\degr\,2046 exhibits only low-significance power ($\log_{10}{\rm FAP}\ge-16$) with semi-amplitudes $\lesssim0.07$\,mmag and no reinforcing ACF/CS signature, consistent with granulation \citep{Kallinger2014} and/or residual \ktwo\ roll-angle systematics \citep{Aigrain2016}. In \tess\ (Sectors\,45 and 72), the star shows a coherent modulation near $P\simeq6.0$--6.4\,d with agreement across the photometric diagnostics (Table\,\ref{tab:var_summary}). Given the small $v\sin i$ (Table\,\ref{tab:sspec_para}), rotational period $P\simeq6$\,d would require a near pole-on inclination. 

TYC\,1395$-$855$-$1 is not detected as variable in \ktwo\ C5 and shows only weak, marginal signals in \ktwo\ C18 ($\log_{10}{\rm FAP}=-9$; Table\,\ref{table:KepWhite2018}), with intermittent WPS power and no coherent ACF signature. In contrast, \tess\ Sector\,72 reveals clear low-frequency variability, including significant power near $P\simeq6.4$\,d and additional components at $P\simeq2.2$--4.4\,d (Table\,\ref{tab:tess}), with concordant photometric diagnostics. For TYC\,1395$-$855$-$1, the small $v\sin i$ (Table\,\ref{tab:sspec_para}) implies that a rotational interpretation would require a near pole-on geometry. 

The discrepancy between the weak or absent \ktwo\ variability and the coherent \tess\ signals in BD\,+19\degr\,2046 and TYC\,1395$-$855$-$1 is significant. Since \tess\ has a much larger pixel scale \citep{2015JATIS...1a4003R} than \ktwo\  \citep{2016ksci.rept....1V}, the extracted \tess\ light curves are more susceptible to flux contamination from nearby unresolved or partially blended sources. Therefore, although the \tess\ signals are coherent and appear significant in the photometric diagnostics, we cannot exclude the possibility that some or all of the detected low-frequency variability originates from contaminating stars within the \tess\ aperture rather than from the target stars themselves. We consequently classify the \tess\ detections in BD\,+19\degr\,2046 and TYC\,1395$-$855$-$1 as low-frequency variability of uncertain origin, pending confirmation from higher-spatial-resolution photometry or time-series spectroscopy.

In summary (see Table\,\ref{tab:var_summary}), the variability of HD\,73135 is best explained by rotational modulation (supported by the $v\sin i$--$R$ constraint in \ktwo) but ellipsoidal variability remains a viable alternative without radial-velocity constraints, BD\,+19\degr\,2045 is a robust new $\gamma$\,Doradus pulsator (rotation ruled out by $P_{\max}$), and BD\,+19\degr\,2046 and TYC\,1395$-$855$-$1 are non-variable or only marginally variable in \ktwo\ but display coherent low-frequency signals in \tess, which we conservatively classify as uncertain pending additional constraints.

\section{Spectral Classification}
\label{sec:class}
Our primary goal is to establish temperature and luminosity anchors for the detailed abundance and forward modelling routine that follows (Sects.\,\ref{para} and \ref{evolv}).  We adopted the Morgan–Keenan (MK) framework, which gives a uniform grid of spectral and luminosity types against which chemical peculiarities can be gauged \citep{1943assw.book.....M, 2009ssc..book.....G}.

Spectra were classified with the automated {\sc Mkclass} code \citep{2014AJ....147...80G}.  The programme implements the MK "metric‑distance" technique \citep{1994ASPC...60..312L}: each continuum‑normalised target spectrum is degraded to the resolution of the reference library (${\cal R}\simeq2500$), rebinned to the 380–462\,nm range, and compared to standard stars through a weighted least‑squares metric.  Separate metric-distance (least-squares mismatch) values are computed for the Ca\,{\sc ii}\,K line (k‑type), the Balmer lines (h‑type), and the metal‑line region (m‑type). If the k-, h-, and m-type classifications are consistent within the adopted uncertainty, we quote a single summary MK type. Otherwise we report the k/h/m types explicitly, since their mismatch is physically informative (e.g. for Am stars).  Our standards are sourced from the Gray–Miller CCD library \citep{2003AJ....126.2048G}; additional plate‑scanned spectra from \citet{1965PASP...77..184C} and \citet{1988A&AS...74..449R} were used as tie‑ins for the late‑K range.

Monte‑Carlo tests with injected noise show that {\sc Mkclass} recovers the correct spectral sub-type  within $\pm1$ subclass and the luminosity class to $\pm0.5$ class for spectra with ${\rm SNR} > 75$, comparable to our HERMES data.  Systematic errors may arise from rotational broadening (\vsini\,$\gtrsim70$\,\kms) which weakens metallic blends and can bias the m‑type to earlier classes; imperfect continuum placement when degrading HERMES ${\cal R}\sim85\,000$ spectra to ${\cal R}=2500$; the limited blue–violet coverage of the standard library, precluding the use of Ca\,{\sc i} 4226\,\AA\ blend often used to separate K\,IV from K\,III.

Accordingly, we quote all the final classifications with a conservative "$\pm1$ subclass, $\pm0.5$ luminosity class" uncertainty.  CN‑strength suffixes (CN1, CN2) are tentative because they rely on relative band‑depths in spectra whose pseudo‑continuum may still retain low‑order blaze residuals.

Table\,\ref{table:class} summarises the classifications. HD\,73135 $\to$ kA7\,hF1\,mF2: the five‑subclass k–metal discrepancy identifies a classical Am star, consistent with the rotational modulation found in Sect.\,\ref{Nat_Var}. BD\,+19\degr\,2045 $\rightarrow$ F8\,V\,($\pm1$): consistent with the photographic F6\,V classification of \citet{1956PASP...68..318B} within the typical $\sim$1--2 subtype uncertainty expected for photographic/low-resolution spectral classifications (mean rms $\simeq1.7$ subtypes; \citealt{1994MNRAS.269...97V}); metal lines appear normal.  BD\,+19\degr\,2046 $\to$ K1\,IV‑V, CN1: the CN enhancement is mild ($1\sigma$ detection) and will be revisited in the abundance analysis. TYC\,1395‑855‑1 $\to$ K2\,IV, CN2: strong CN bands suggest dredge‑up of CNO‑processed material, again to be checked in Sect.\,\ref{para}.

\begin{table}
\centering
\caption{ Spectral classifications (typical uncertainty: $\pm1$ subclass, $\pm0.5$ luminosity class). A single MK type is quoted when k/h/m diagnostics agree; otherwise the k/h/m types are reported.}
\label{table:class}
\begin{tabular}{lcc}
\hline
Star & This work & Literature \\
\hline
HD\,73135       & kA7\,hF1\,mF2 (Am)  & A2\,(1), Am\,(2) \\
BD\,+19\degr\,2045   & F8\,V                & F6\,V\,(3) \\
BD\,+19\degr\,2046   & K1\,IV–V, CN1        & K0\,(4) \\
TYC\,1395‑855‑1  & K2\,IV, CN2          & — \\
\hline
\end{tabular}
\\[2pt]
\textbf{References:} (1) \citet{1993yCat.3135....0C};
(2) \citet{1965PASP...77..184C}; (3) \citet{1956PASP...68..318B};
(4) \citet{1988A&AS...74..449R}
\end{table}

These spectral types provide the backbone for the atmospheric‑parameter and evolutionary modelling that follow in Sects.\,\ref{para} and\,\ref{evolv}.

\section{Astrophysical Parameters and Chemical Abundances}
\label{para}

In order to obtain a self‑consistent set of atmospheric parameters, we combined three complementary techniques, intermediate‑band photometry, broad‑band spectral‑energy‑distribution (SED) fitting, and high‑resolution synthetic‑spectrum fitting. Photometry and SED fitting provide initial estimates of stellar parameters, which are then refined through spectroscopic analysis and subsequently used to determine individual chemical abundances.  
The steps needed to estimate basic physical parameters, as well as the main uncertainties, are summarised here.
\subsection{Photometry}
\label{photo}

Str\"{o}mgren $uvby\beta$ indices were chosen because the combination of their $m_{1}$ and $c_{1}$ indices clearly separate effective temperature, surface gravity, and metallicity for A--G stars \citep{1985MNRAS.217..305M}. We adopted the \citet{1985MNRAS.217..305M} calibrations to derive photometric values of \teff, \logg, and [M/H]. The uncertainties quoted for the $uvby\beta$-based parameters listed in Table\,\ref{table:basic} (i.e.\ $\pm200$\,K in \teff, $\pm0.10$\,dex in \logg, and $\pm0.13$\,dex in [M/H]) are adopted errors that incorporate measurement uncertainties, reddening uncertainties, and calibration scatter. The reddening $E{\rm (B-V)}$ was estimated from the 3-D dust map of \citet{bayestar} and \citet{2019ApJ...887...93G}.

Near-infrared 2MASS colours anchor the Rayleigh--Jeans tail and help break the reddening--temperature degeneracy. The 2MASS $JHK_s$ photometry was taken from the 2MASS All-Sky Point Source Catalog \citep{2003yCat.2246....0C}. We used the catalogue magnitudes and uncertainties to form near-IR colours (e.g. $J-K_s$), which were dereddened using the adopted $E{\rm (B-V)}$ and a standard extinction law \citep[e.g.][]{1985ApJ...288..618R, 1989ApJ...345..245C}. Effective temperatures were then estimated from $(J-K_s)_0$ using the InfraRed Flux Method (IRFM) colour--$T_{\rm eff}$ calibration of \citet{2009A&A...497..497G}. Since the 2MASS-based temperatures are used only as initial guesses for the spectroscopic analysis, we assume solar metallicity in the calibration. The quoted 2MASS-based \teff\ uncertainties listed in Table\,\ref{table:basic} include propagated photometric and reddening errors together with the intrinsic scatter of the calibration.


\subsection{Effective temperature and \logg\ from SED fitting}
\label{sed}

SEDs were built with {\sc VOSA} \citep{2008A&A...492..277B} using photometric measurements extracted from the Tycho-2, 2MASS, WISE, \textit{Gaia}\,eDR3, and (when available) GALEX, SDSS, and APASS databases. Kurucz ATLAS9 ODFNEW/NOVER models were fitted via $\chi^{2}$ minimisation, using \textit{Gaia} parallaxes as distance priors and $A_{V}=3.1\,E{\rm(}B-V{\rm)}$ extinction.  We fixed the model metallicity to solar composition ([M/H]$=0.0$) in the SED fitting \citep[e.g.][]{10.1093/mnras/sty722, 2019MNRAS.484.2530C}. Formal {\sc VOSA} errors underestimate true systematics, so we follow the technique described by \citet{2021MNRAS.504.5528T} and adopt $\sigma_{T_{\rm eff}}=250$\,K, $\sigma_{\log g}=0.25$\,dex.

\begin{table*}
 \centering
\caption{Basic parameters of the sample.   Right ascension and declination (J2000) are from Gaia DR3 \citep{Gaia3}.}

\begin{tabular}{lcccccccccccc}
 \hline\hline
 \noalign{\smallskip}
 Star&$\alpha$ (J2000)&$\delta$ (J2000)& $E{\rm (B-V)}$&$T_{\rm eff}$&\logg&[M/H]&$T_{\rm eff}$&$T_{\rm eff}$&\logg\\
 &&&&${}^{uvby\beta}$&${}^{uvby\beta}$&${}^{uvby\beta}$& ${}^{\rm 2MASS}$&${}^{\rm SED}$&${}^{\rm SED}$\\
 ID& (hh mm ss) &(dd mm ss) &(mag)&($\pm 200$ K)&($\pm 0.10$\,dex)&$\pm 0.13$&(K)&($\pm 250$ K)&($\pm 0.25$\,dex)\\
 \noalign{\smallskip}
 \hline
 HD\,73135&08 37 18.2273&+18 52 11.8914&0.010\,$\pm$\,0.002&-&-&-&7360$\pm$250& 7250& 4.0\\
 BD\,$+19^{\circ}2045$&08 36 29.8499&+18 57 57.0725&0.011\,$\pm$\,0.002&6580&4.10&0.15&6480$\pm$220&6250& 4.0\\
 BD\,$+19^{\circ}2046$&08 36 34.5373& +18 51 13.2893&0.026\,$\pm$\,0.005&-&-&-&4770$\pm$190&4750&3.0\\
 TYC\,1395-855-1&08 37 01.6850& +19 01 38.6082&0.032\,$\pm$\,0.006&-&-&-&5040$\pm$210&4750&3.25\\
 \noalign{\smallskip}
\hline\\
 \end{tabular}
\label{table:basic}
\end{table*}

\subsection{Radial velocity}
\label{sect:vrad}

Absolute velocities were measured with the Fourier cross‑correlation algorithm of \citet{1979AJ.....84.1511T} as implemented in {\sc iSpec} \citep{2014ASInC..11...85B, 2019arXiv190209558B}. To ensure line‑list consistency, we used synthetic templates computed at the \teff\ and \logg\ obtained from our photometry and SED analyses (Sects.\,\ref{photo} and \ref{sed}). A solar metallicity was assumed because \vrad\ is  insensitive to modest abundance errors. The results are presented in Table\,\ref{tab:sspec_para}. Statistical errors come from the Tonry–Davis $R_{\rm TD}$ value; a systematic $\pm0.2$\,\kms\ floor (barycentric correction and instrumental drift) is added quadratically.

\subsection{Projected rotational velocity}
\label{sect:vsini}

Projected rotational velocity (\vsini) was derived entirely from synthetic-profile fitting.  For each star, we generated a local ATLAS9 synthetic spectrum (the initial guesses of \teff\ and \logg, solar abundances, and \vrad\ as in Sect.\,\ref{sect:vrad}), convolved it with a rotational broadening kernel and performed a least‑squares optimisation with the {\sc Minuit} engine inside {\sc girfit} \citep{2006A&A...451.1053F}.  The fit was repeated for contiguous 10\,nm windows across 500–600\,nm, a region rich in moderately strong metal lines but largely free of Balmer‑wings and telluric contamination. The Mg\,{\sc i} triplet was excluded for BD\,+19\degr\,2046 and TYC\,1395‑855‑1 because its strong damping wings bias the velocity kernel. The \vsini\ grid spanned 0–200\,\kms\ in 1\,\kms\ steps; the segment‑to‑segment scatter is adopted as the total uncertainty.  The final values are listed in Table\,\ref{tab:sspec_para}.

\subsection{\texorpdfstring{\teff\ and \logg\ from spectroscopy}{Teff and log g from spectroscopy}}
\label{spec:teff}

We determined the atmospheric parameters by $\chi^{2}$–minimisation between normalised observed spectra and synthetic calculations. The latter were computed with the \texttt{SPECTRUM} radiative–transfer code, using atomic data from VALD, solar abundances from \citet{2009ARA&A..47..481A}, and plane–parallel LTE ATLAS9 model atmospheres \citep{2003IAUS..210P.A20C}\footnote{\url{http://www.stsci.edu/hst/observatory/crds/castelli_kurucz_atlas.html}}. The workflow was implemented with \textsc{iSpec} \citep{2014ASInC..11...85B,2019arXiv190209558B}.

For the hotter stars (HD\,73135 and BD\,+19\degr\,2045), we used the H$\beta$ wings to refine \teff; and we did not use them to constrain \logg\ \citep{2005MSAIS...8..130S}. Surface gravities were instead determined from fitting the Fe\,\textsc{i}/Fe\,\textsc{ii} lines and we selected only the lines that are unblended at the HERMES resolution and have a well-defined local continuum. For the cooler stars (BD\,+19\degr\,2046 and TYC\,1395$-$855$-$1), where Balmer profiles are weak and more susceptible to systematic effects, \teff\ and \logg\ were determined entirely from fitting Fe\,\textsc{i}/Fe\,\textsc{ii} lines using the same line-selection criteria.

Parameter uncertainties combine the internal errors from the covariance matrix of the fit with systematic terms. Continuum normalisation is a leading source of systematics in high-resolution work due to noise correlations (e.g. rebinning), line crowding, and the need to model broad spectral regions \citep[e.g.][]{Napiwotzki1993,Morel2006}. We estimate the continuum-placement error as:
\begin{equation}
  \epsilon_{\rm cont} \simeq \frac{1.5}{\mathrm{SNR}},
\end{equation}
with SNR measured per pixel \citep{2005oasp.book.....G}. To propagate this term, the observed spectrum was renormalised by factors $(1\pm\epsilon_{\rm cont})$ and the fit repeated; the deviations from the nominal solution were adopted as the continuum-related uncertainties. To account for residual systematics not captured by the internal fit errors and the continuum-renormalisation tests, we impose conservative minimum (floor) uncertainties of 150\,K in \teff\ and 0.15\,dex in \logg\  \citep{2014ApJS..211....2H, 2014MNRAS.440.2665R}, added in quadrature. The uncertainties reported in Table\,\ref{tab:sspec_para} reflect the quadrature sum of the internal errors, the continuum-related terms, and these adopted floors.


Overall, the photometric, SED, and spectroscopic parameter estimates are broadly consistent (Tables\,\ref{table:basic} and \ref{tab:sspec_para}). For HD\,73135, the 2MASS-, SED-, and spectroscopic temperatures agree within $\sim$150\,K; BD\,+19\degr\,2045 likewise shows good consistency, with the spectroscopic solution lying between the photometric and SED estimates. TYC\,1395$-$855$-$1 also shows satisfactory agreement within the adopted uncertainties. The largest discrepancy is found for BD\,+19\degr\,2046, for which the spectroscopic solution is hotter and yields a somewhat higher \logg\ than the photometric/SED estimates. In the analyses that follow, we adopt the spectroscopic parameters as the final values, while the photometric and SED determinations are treated as independent checks and initial constraints.

\subsection{Microturbulent velocity}
\label{sect:micro}

The microturbulent velocity ($\xi$) was determined within the global synthetic-spectrum fit. We constrained $\xi$ by fitting the profiles of selected clean, moderately strong lines, requiring the synthetic spectra to reproduce the relative core and wing strengths. Fe\,\textsc{i} lines provided the primary constraint, while Ti\,\textsc{ii} lines were included only as additional clean features to improve the stability of the fit. We retained only lines that are unblended at the HERMES resolution and have reliable VALD3 atomic data ($\log gf$ uncertainties $\leq0.05$\,dex).

Uncertainties in $\xi$ were estimated from the covariance matrix of the fit and verified with Monte Carlo noise tests; they scale approximately with $1/\mathrm{SNR}$ and the inverse square root of the number of fitted lines. The resulting 1$\sigma$ uncertainties are typically 0.21--0.34\,\kms. The final values are listed in Table\,\ref{tab:sspec_para}.

\begin{figure*}

\includegraphics[width=\textwidth]{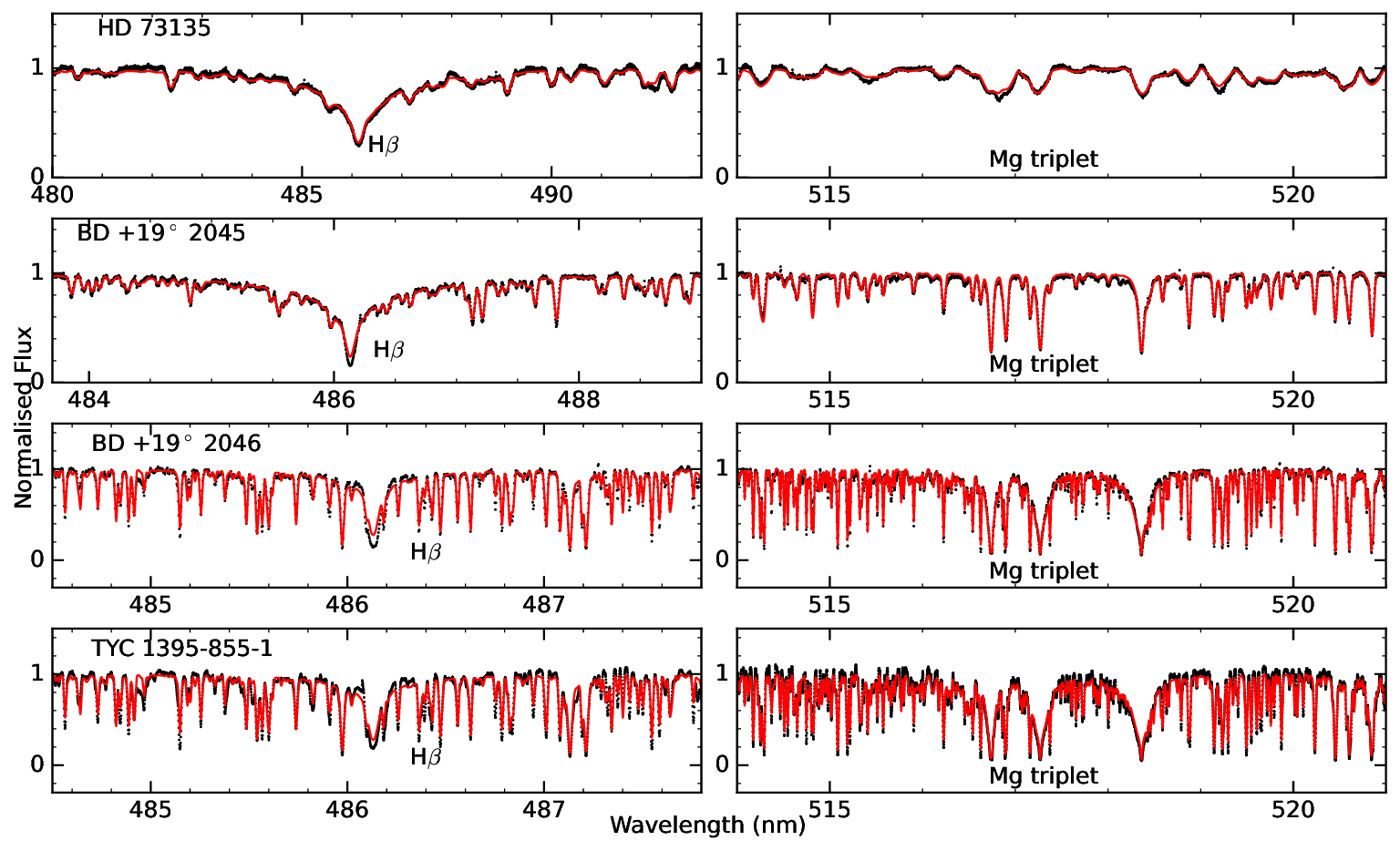}
\caption{The H$\beta$ (left) and Mg\,{\sc i} triplet (right) regions of the spectra for the target stars. In each panel, the observed spectrum (black) is overplotted with the best-fitting synthetic spectrum (red). All spectra have been shifted to the stellar rest frame using the measured radial velocities (Table\,\ref{tab:sspec_para}), so the Doppler shifts are removed by construction.  For the K-type stars, the poorer H$\beta$ fits are expected because the Balmer lines are intrinsically weaker and more affected by metal-line blending and continuum-normalisation uncertainties in this regime. We therefore do not use the Balmer profiles of BD\,+19\degr\,2046 and TYC\,1395$-$855$-$1 as quantitative constraints on the atmospheric parameters.}

\label{sphd73135}
\end{figure*}

\begin{table*}
\centering
\caption{The fundamental astrophysical parameters as computed from high-resolution spectroscopy and the log$\left(L_\star/L_\odot\right)$ as calculated from the GAIA DR3 parallaxes are listed. Quoted uncertainties include internal fit errors, continuum-renormalisation tests, and adopted systematic floors added in quadrature (Sect.\,\ref{spec:teff}).}
\label{tab:sspec_para}
\begin{tabular}{lccccccc}
\hline\hline
Star  & $T_{\rm eff}$&\logg& [M/H]$_{\rm model}$&[Fe/H] & $\xi$& \vsini & $V_{\rm rad}$ \\
ID& (K)&(dex)& (dex)&(dex)&(\kms) &(\kms) & \\
\hline
HD\,73135
& 7210\,$\pm$\,190& 3.95\,$\pm$\,0.17& $-0.47\,\pm$\,0.08& $-0.08\,\pm$\,0.06 & 3.48\,$\pm$\,0.34 & 66.0\,$\pm$\,2.8 & 10.22\,$\pm$\,1.46 \\

BD\,+19\degr\,2045 & 6440\,$\pm$\,230 & 4.05\,$\pm$\,0.20 & $-0.09\,\pm$\,0.11 & $+0.04\,\pm$\,0.06
& 2.05\,$\pm$\,0.25 & 17.0\,$\pm$\,2.2 & 38.10\,$\pm$\,0.22 \\

BD\,+19\degr\,2046 & 5270\,$\pm$\,190 & 3.45\,$\pm$\,0.23 & $-0.13\,\pm$\,0.06 & $-0.08\,\pm$\,0.05
& 1.37\,$\pm$\,0.23 & 6.0\,$\pm$\,2.8 & 45.65\,$\pm$\,0.20 \\

TYC\,1395-855-1 & 5000\,$\pm$\,210 & 3.13\,$\pm$\,0.27 & $-0.04\,\pm$\,0.05 & $+0.04\,\pm$\,0.05
& 1.10\,$\pm$\,0.21 & 5.0\,$\pm$\,2.2 & $13.78\,\pm$\,0.20 \\
\hline
\end{tabular}
\end{table*}

\subsection{Individual Chemical Abundances}
\label{abundances}

The chemical abundance analysis was carried out using the spectroscopically determined fundamental parameters as input. Each stellar spectrum was partitioned into several local fitting windows, typically spanning 5--10\,nm (50--100\,\AA), to enable reliable local continuum placement and line-profile fitting. These windows were selected manually, centred on diagnostically useful lines, and chosen to avoid severe blends, continuum defects, and telluric contamination. Within each window, elemental abundances were derived through a least-squares minimisation procedure \citep{2007nure.book.....P}.

We used the \textsc{SynthV\_NLTE} spectral synthesis code \citep{2019ASPC..518..247T} alongside a grid of stellar atmosphere models precomputed with the \textsc{LLmodels} code \citep{2004A&A...428..993S}. The synthesis process was facilitated using the IDL-based interface \textsc{BinMag6} \citep{2018ascl.soft05015K}. All calculations were performed under the assumption of local thermodynamic equilibrium (LTE). Atomic data and spectral line parameters were adopted from the third release of the Vienna Atomic Line Database (VALD3; \citealt{ryabchikova2015}).

\begin{figure}
\centering
 \includegraphics[width=\columnwidth]{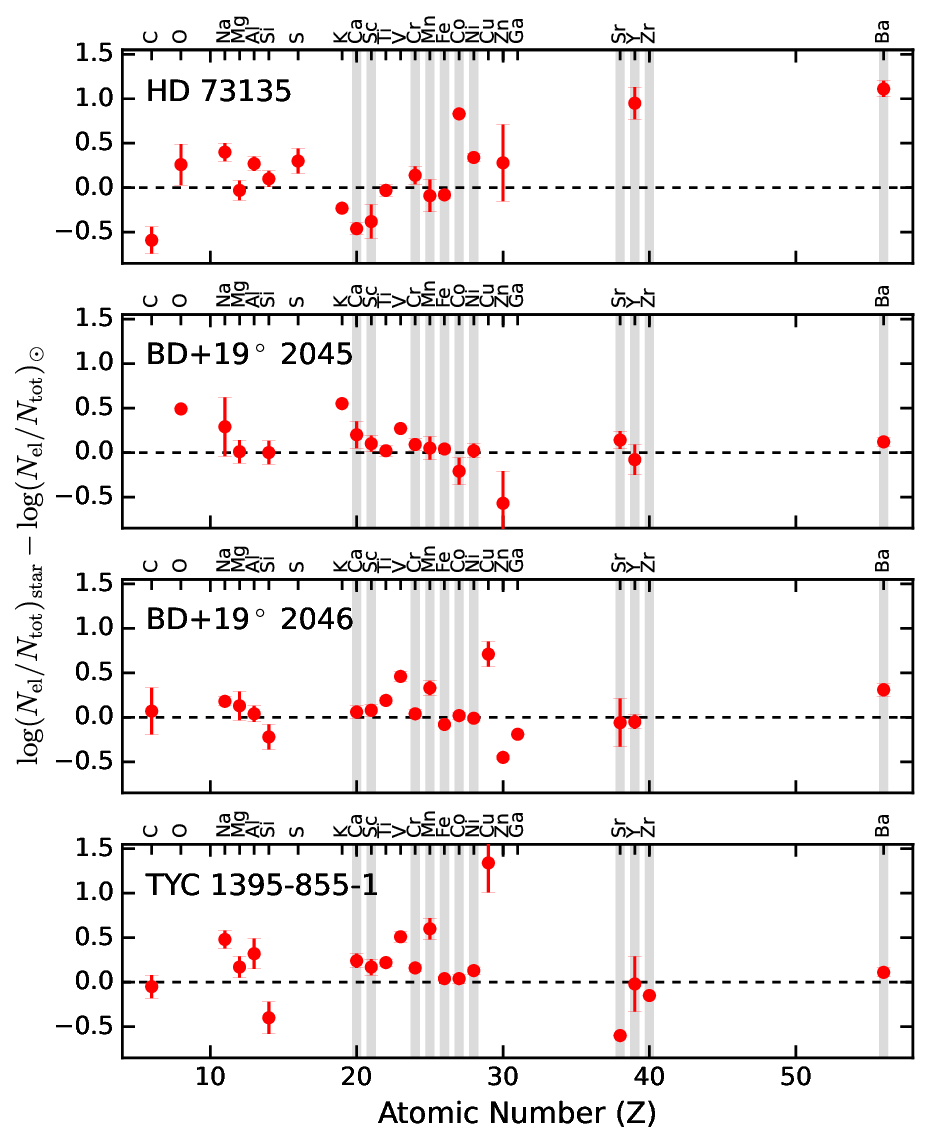}
\caption{The abundances of chemical elements of the target stars as a function of atomic number relative to the solar values by \citet{2009ARA&A..47..481A}. The grey vertical background bars highlight the light elements (Ca and Sc) and heavy elements (Cr, Mn, Fe, Co, Ni, Sr, Y, Zr, and Ba) relevant for Am star classification.}
\label{abundance:fig}
\end{figure}

For each chemical species, the final abundance was calculated as the mean of the values obtained from all relevant spectral segments. The standard error of the mean was taken as the uncertainty when an element was detected in multiple segments. Table\,\ref{table_abundances} lists the measured abundances along with the number of contributing segments (in parentheses). The overall abundance patterns of the analysed stars, referenced to the solar composition of \citet{2009ARA&A..47..481A}, are illustrated in Fig.\,\ref{abundance:fig}.

HD\,73135 exhibits marked chemical peculiarities characteristic of an Am star, with strong overabundances of Sr and Ba ([X/H]\,$\gtrsim$\,+1) together with clear underabundances of Ca and Sc. This combination of heavy-element enhancement and Ca--Sc depletion is the defining abundance signature of the chemically peculiar object in our sample.

BD\,+19\degr\,2045 does not exhibit the characteristic abundance pattern of an Am star. In contrast to HD\,73135, it shows neither the coupled Ca--Sc depletion nor the pronounced enhancement of iron-peak and heavy elements that typify metallic-line A stars. Its abundance distribution is therefore consistent with a chemically normal late-F star, within the measurement uncertainties and the expected element-to-element scatter. We consequently classify BD\,+19\degr\,2045 as chemically non-peculiar. 

The two K-type stars, BD\,+19\degr\,2046 and TYC\,1395$-$855$-$1, are discussed only in terms of their general photospheric composition. 
BD\,+19\degr\,2046 is approximately solar in metallicity, with $\Delta \log N({\rm Fe}) \simeq -0.08$ dex, and most elements lying within a few tenths of a dex of the solar reference. Moderate deviations are seen for some species, including enhancements of Ti, V, Mn, Cu, and Ba, and subsolar Si, Zn, and Ga. TYC\,1395$-$855$-$1 is also close to solar in iron, with $\Delta \log N({\rm Fe}) \simeq +0.04$ dex, but shows mild enhancements in several light and iron-peak elements, particularly Na, Mg, Al, Ti, V, Mn, and Cu, while Si and Sr are subsolar. These element-to-element variations are treated as part of the general spectroscopic characterization of the stars.

The pronounced heavy-element enhancements together with Ca--Sc depletion in HD\,73135 are consistent with radiative diffusion and gravitational settling in a stable stellar atmosphere. By contrast, the other three stars show no comparable, internally consistent Am-type pattern. For chemically peculiar stars, a single scaled-solar metallicity parameter [M/H] (as used in global spectral fitting) should not be interpreted as a true bulk metallicity; the chemically meaningful description is instead provided by [Fe/H] (Table\,\ref{tab:sspec_para}) and the element-by-element abundances (Table\,\ref{table_abundances}).

\subsubsection{NLTE Effects on Derived Abundances}

A brief assessment of NLTE effects indicates that HD\,73135 (\teff\,$=7210\pm110\,$K, $\log g=3.95\pm0.08$ (dex)) requires the largest abundance adjustments up to $\sim$0.2\,dex for the O\,\textsc{i}\,777\,nm triplet, Na\,\textsc{i} D\,lines, and Fe\,\textsc{i}, while its Sr\,\textsc{ii} and Ba\,\textsc{ii} resonance lines incur corrections of $\lesssim$0.1\,dex \citep{2009A&A...494.1083A, 2011A&A...528A.103L, 2011A&A...528A..87M,2013AstL...39..126S}.  For BD\,+19\degr\,2045 (\teff\,$=6440\pm180\,$K, \logg\,$=4.05\pm0.13$ (dex)), offsets shrink to $\sim$0.05–0.08\,dex (Na, Mg) and $\leq$0.05\,dex for Fe\,\textsc{i} and heavy‐element lines \citep{2012MNRAS.427...27B}.  The cooler stars BD\,+19\degr\,2046 and  TYC\,1395‑855‑1 exhibit NLTE corrections $\lesssim$0.05\,dex for all species \citep{2007A&A...461..261M}.  Thus, while LTE abundances suffice to better than 0.05\,dex for the cooler targets, line‐by‐line NLTE corrections are advisable for HD\,73135 when pursuing precision below this threshold. 

\smallskip
\noindent
With these methodological clarifications, the parameters in Table\,\ref{tab:sspec_para} and the abundances in Table\,\ref{table_abundances} can be considered internally consistent and sufficiently accurate for the evolutionary analysis in Sect.\,\ref{evolv}.

\begin{table*}
\centering
\caption{Individual chemical abundances inferred for HD\,73135, BD\,+19\degr\,2045, BD\,+19\degr\,2046, and  TYC\,1395-855-1. The solar abundances are listed in the last column \citep{2009ARA&A..47..481A}. The values are expressed in the form log$(N_{\rm el}/N_{\rm Tot})$. The number of segments from which each element was determined is given in brackets.}
\label{table_abundances}
\begin{tabular}{lccccr}
\hline\hline
\noalign{\smallskip}
Element	&HD\,73135 &BD\,+19\degr\,2045 &BD\,+19\degr\,2046 & TYC\,1395-855-1 & Solar Values\\
\noalign{\smallskip}
\hline
 C & -4.16\,$\pm$\,0.15 (11) & - & -3.50\,$\pm$\,0.26 (11) & -3.62\,$\pm$\,0.13 (3) & -3.57\\
  O & -3.05\,$\pm$\,0.23 (2) & -2.82 (1) &  - &  - & -3.31\\
  Na & -5.36\,$\pm$\,0.10 (3) & -5.47\,$\pm$\,0.33 (4) & -5.58\,$\pm$\,0.05 (3) & -5.28\,$\pm$\,0.10 (7) & -5.76\\
  Mg & -4.43 \,$\pm$\, 0.11 (12) & -4.39\,$\pm$\,0.13 (7) & -4.27\,$\pm$\,0.16 (13) & -4.23\,$\pm$\,0.12 (10) & -4.4\\
  Al & -5.28 \,$\pm$\, 0.08 (4) & -  & -5.51\,$\pm$\,0.09 (6) & -5.23\,$\pm$\,0.17 (2) & -5.55\\
  Si & -4.39 \,$\pm$\, 0.09 (18) & -4.49\,$\pm$\,0.13 (14) & -4.71\,$\pm$\,0.14 (12) & -4.89\,$\pm$\,0.18 (5) & -4.49\\
  S & -4.58 \,$\pm$\, 0.14 (7) & -  & -  & - & -4.88\\
  K & -7.20 (1) & -6.42 (1) & -   & - & -6.97\\
  Ca & -6.12 \,$\pm$\, 0.07 (19) & -5.46\,$\pm$\,0.15 (17) & -5.60\,$\pm$\,0.07 (26) & -5.42\,$\pm$\,0.08 (19) & -5.66\\
  Sc & -9.23 \,$\pm$\, 0.19 (11) & -8.75\,$\pm$\,0.09 (9) & -8.77\,$\pm$\,0.07 (17) & -8.68\,$\pm$\,0.09 (12) & -8.85\\
  Ti & -7.08 \,$\pm$\, 0.07 (28) & -7.03\,$\pm$\,0.06 (22) & -6.86\,$\pm$\,0.03 (32) & -6.83\,$\pm$\,0.04 (27) & -7.05\\
  V &  -  & -7.80 (1) & -7.61\,$\pm$\,0.06 (17) & -7.56\,$\pm$\,0.06 (17) & -8.07\\
  Cr & -6.22 \,$\pm$\, 0.10 (20) & -6.27\,$\pm$\,0.07 (13) & -6.32\,$\pm$\,0.05 (23) & -6.20\,$\pm$\,0.04 (20) & -6.36\\
  Mn & -6.66 \,$\pm$\, 0.18 (4) & -6.52\,$\pm$\,0.13 (6) & -6.24\,$\pm$\,0.08 (16) & -5.97\,$\pm$\,0.12 (15) & -6.57\\
  Fe & -4.58 \,$\pm$\, 0.04 (51) & -4.46\,$\pm$\,0.04 (41) & -4.58\,$\pm$\,0.02 (37) & -4.46\,$\pm$\,0.02 (33) & -4.5\\
  Co & -6.18 (1) & -7.22\,$\pm$\,0.15 (2) & -6.99\,$\pm$\,0.04 (23) & -6.97\,$\pm$\,0.05 (17) & -7.01\\
  Ni & -5.44\,$\pm$\,0.04 (33) & -5.76\,$\pm$\,0.08 (20) & -5.79\,$\pm$\,0.02 (27) & -5.65\,$\pm$\,0.04 (27) & -5.78\\
  Cu &   - &  - & -7.10\,$\pm$\,0.14 (2) & -6.47\,$\pm$\,0.33 (3) & -7.81\\
  Zn & -7.16 \,$\pm$\, 0.43 (4) & -8.01\,$\pm$\,0.36 (2) & -7.89\,$\pm$\,0.02 (2) & - & -7.44\\
  Ga &   - & -  & -9.15 (1) & -  & -8.96\\
  Sr &  -  & -8.99\,$\pm$\,0.10 (3) & -9.19\,$\pm$\,0.27 (3) & -9.73 (1) & -9.13\\
  Y & -8.84 \,$\pm$\,0.18 (8) & -9.87\,$\pm$\,0.17 (8) & -9.84\,$\pm$\,0.08 (9) & -9.81\,$\pm$\,0.31 (5) & -9.79\\
  Zr &   - &  - &  - & -9.57 (1) & -9.42\\
  Ba & -8.71\,$\pm$\,0.09 (6) & -9.70\,$\pm$\,0.05 (6) & -9.51\,$\pm$\,0.07 (6) & -9.71\,$\pm$\,0.05 (7) & -9.82\\
\noalign{\smallskip}
\hline\\
\end{tabular}
\end{table*}

\section{Fundamental Parameters and Evolutionary Phases}
\label{evolv}

\subsection{Masses, radii, and ages from forward modelling}
\label{sect:param}
We adopt the MIST stellar evolutionary tracks (\citealp[]{Choi_2016ApJ...823..102C,Dotter_2016ApJS..222....8D}) which are computed using a  one-dimensional (1D) stellar evolution code MESA (Modules for Experiments in Stellar Astrophysics; \citealt{Paxton_2011ApJS..192....3P,Paxton_2013ApJS..208....4P,   Paxton_2015ApJS..220...15P, Pax4, 2019ApJS..243...10P, 2023ApJS..265...15J}). We used MIST tracks for three reasons: (i) MIST is computed with a single version of the MESA code, ensuring that low‑ and intermediate‑mass tracks share identical opacities, nuclear rates, overshoot and diffusion prescriptions, important when comparing an Am star, an F dwarf, and K-subgiants within one sample. (ii)  The grid spans $-2.0\le[\mathrm{Fe/H}]\le+0.5$\,dex in 0.25‑dex steps, fully bracketing our spectroscopic metallicities. (iii) Fine timesteps ($\Delta\log t=0.02$) minimise interpolation artefacts in the subgiant phase, critical for modelling BD\,+19\degr\,2046 and TYC\,1395‑855‑1.

For each star, we take the spectroscopic
$T_{\rm eff}$, $\log(L_\star/L_\odot)$, $\log g$, and [Fe/H] computed as described in Sect.\,\ref{para}, and compare them with their corresponding model quantities  $ T_{\rm eff,mod}$, $\log L_{\rm mod}$, $\log g_{\rm mod}$, and [Fe/H]$_{\rm mod}$ using a $\chi^{2}$ expressed as: \begin{equation} 
\chi^2 = \sum_i
         \frac{({\rm obs}_i-{\rm mod}_i)^2}{\sigma_i^2},
\end{equation}
 where ${i\in\{T_{\rm eff},\log(L_\star/L_\odot),\log g,[{\rm Fe/H}]\}}$.
 
We note that stellar luminosities, expressed as $\log(L_\star/L_\odot)$, were determined using Gaia\,DR3 parallaxes (including zero‐point corrections; \citealt{2023A&A...674A..32B}) and apparent magnitudes, with interstellar extinction $A_V$ as described in Sect.\,\ref{sed}.  Bolometric corrections were adopted from \citet{1996ApJ...469..355F}, updated by \citet{2010AJ....140.1158T}.  The resulting $\log(L_\star/L_\odot)$ values for all programme stars are given in Table\,\ref{tab:param_results}.

For each MIST grid point, we compute a relative posterior weight:
\begin{equation}
w \propto e^{-\chi^{2}/2}\,\Delta t,
\end{equation}
where $e^{-\chi^{2}/2}$ is the likelihood term and $\Delta t$ is the local timestep spacing in linear age, included to account for the evolutionary dwell time. These weights were calculated in our forward-modelling routine, normalised over the grid, and then used to construct the weighted distributions of mass, radius, and age. The $16^{th}$, $50^{th}$, and $84^{th}$ percentiles of these distributions provide the median estimates and associated 68\% credible intervals; these results are presented in Table\,\ref{tab:param_results}.

\begin{table}
  \centering
  \caption{Fundamental parameters for the programme stars derived from MIST.  Uncertainties denote the 16th–84th percentile range of the posterior.}
  \label{tab:param_results}
  \begin{tabular}{lcccc}
    \hline
    Star & $\log(L_\star/L_\odot)$ & $M$ (M$_\odot$) & $R$ (R$_\odot$) & Age (Gyr) \\
    \hline
    HD\,73135 &1.16$\pm0.05$      & \(1.76^{+0.04}_{-0.05}\) & \(2.40^{+0.15}_{-0.15}\) & \(1.12^{+0.14}_{-0.00}\) \\
    BD\,+19\degr\,2045 &0.75$\pm$0.18    & \(1.32^{+0.09}_{-0.07}\) & \(1.91^{+0.23}_{-0.21}\) & \(3.16^{+0.82}_{-0.65}\) \\
    BD\,+19\degr\,2046 & 1.48$\pm$0.06    & \(2.25^{+0.10}_{-0.09}\) & \(6.33^{+0.59}_{-0.51}\) & \(0.79^{+0.10}_{-0.09}\) \\
    TYC\,1395$-$855$-$1 & 1.16$\pm0.06$ &  \(1.38^{+0.31}_{-0.25}\) & \(5.33^{+0.42}_{-0.41}\) & \(3.55^{+3.53}_{-1.77}\) \\
    \hline
  \end{tabular}
\end{table}

\begin{figure}
\centering
\includegraphics[width=\columnwidth]{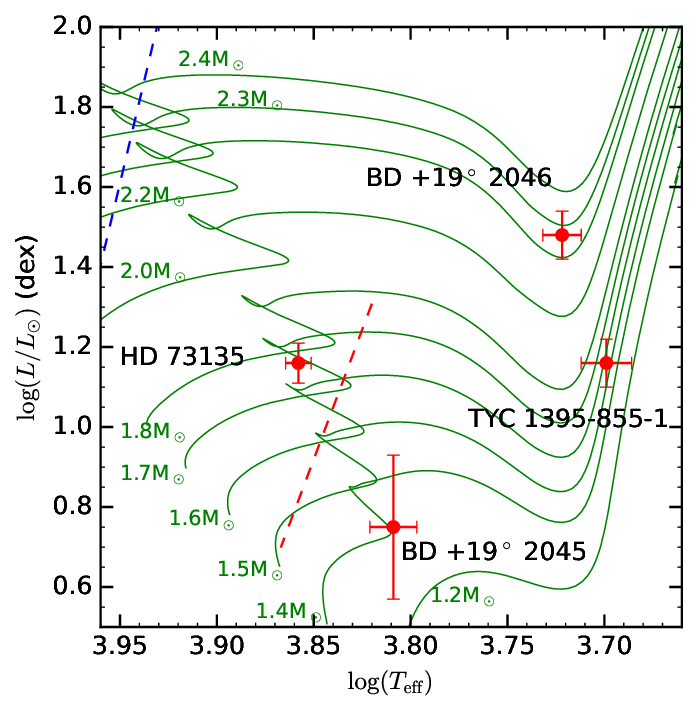}
\caption{ Hertzsprung–Russell diagram showing the positions of the sample stars. Solid green lines represent MIST evolutionary tracks for masses between 1.2\,M$_\odot$ and 2.4\,M$_\odot$. Blue and red dashed lines mark the empirical boundaries of the $\delta$\,Scuti instability strip, as defined by \citet{10.1093/mnras/stz590}.}
\label{fig:HR}
\end{figure}

To assess the evolutionary status, we plot our target stars on the Hertzsprung–Russell diagram (Fig.\,\ref{fig:HR}), alongside 
evolutionary tracks (1.2–2.4\,M$_\odot$ in 0.2\,M$_\odot$ steps).  HD\,73135 and BD\,+19\,2045 lie near the terminal‐age main sequence, BD\,+19\,2046 is located at the base of the red giant branch, and TYC\,1395‑855‑1 is ascending the red giant branch.  HD\,73135 falls within the $\delta$\,Scuti instability strip \citep{10.1093/mnras/stz590}, consistent with its classification as a potential pulsator.

\section{Conclusions}
\label{conclusion}

We have presented a combined analysis of ground-based, \ktwo, and \tess\ photometry together with high-resolution HERMES spectroscopy for four bright stars (HD\,73135, BD\,+19\degr\,2045, BD\,+19\degr\,2046, and TYC\,1395-855-1) originally monitored, but not fully characterised, under the Nainital--Cape Survey.   We aimed to (i) search for and characterise low-amplitude variability and assess the consistency of multiple photometric diagnostics, (ii) determine spectral classifications and atmospheric parameters and identify chemical peculiarities where present, and (iii) infer fundamental parameters, including mass, radius, and age, using MESA Isochrones and Stellar Tracks (MIST) evolutionary models.

The ground-based Johnson $B,V$ photometry does not reveal any significant intrinsic periodicity, and the strongest frequency peaks are consistent with diurnal window-function and extinction-related artefacts. By contrast, the space-based photometry reveals clear low-amplitude variability, but with different behaviours in the four stars. HD\,73135 shows a persistent $\sim$1.5\,d signal in \ktwo, which is most plausibly interpreted as rotational modulation, although an ellipsoidal-binary interpretation cannot yet be excluded without time-series radial velocities. BD\,+19\degr\,2045 shows a stable pattern of low-frequency variability consistent with a candidate $\gamma$\,Doradus pulsator. BD\,+19\degr\,2046 and TYC\,1395-855-1 are non-variable or only marginally variable in \ktwo, but both show coherent low-frequency modulation in \tess, which we conservatively classify as variability of uncertain origin.

The spectroscopic analysis identifies HD\,73135 as the only chemically peculiar object in the sample, with an Am-star abundance pattern, while BD\,+19\degr\,2045 is a chemically normal F8\,V star and BD\,+19\degr\,2046 and TYC\,1395-855-1 are chemically non-peculiar K-type subgiants. Forward modelling places HD\,73135 and BD\,+19\degr\,2045 near the main sequence whereas the two K stars are at more evolved subgiant stages.

The combination of time-series photometric observations with longer-baseline multi-epoch spectroscopy will be essential for refining the variability classifications and testing the physical origin of the signals in all studied stars. The upcoming space mission, \plato\ \citep{2025ExA....59...26R}, will also provide valuable insights for improving the photometric characterisation of BD\,+19\degr\,2046 and TYC\,1395-855-1. 


\section*{Data availability statement}

The high-resolution spectroscopic data used in this article will be shared upon reasonable request. The \ktwo\ and \tess\ light curves for the target stars are publicly available on the MAST archive\footnote{https://mast.stsci.edu/portal/Mashup/Clients/Mast/Portal.html}.

\section*{Acknowledgments}
We acknowledge funding through the Max Planck – Humboldt Research Unit, a partnership between the Max Planck Institute for Solar System Research (MPS) and Kyambogo University (KyU). BN also acknowledges additional funding through the Max Planck Partnership Group – SEISMIC project, a collaboration between the Max Planck Institute for Astrophysics (MPA), Germany, and Kyambogo University (KyU), Uganda. OT and EJ thank the International Science Programme (ISP) of Uppsala University for  financial support. Part of work presented here is supported by the Belgo-Indian Network for Astronomy \& Astrophysics (BINA), approved by the International Division, Department of Science and Technology (DST, Govt. of India; DST/INT/Belg/P-02) and the Belgian Federal Science Policy Office (BELSPO, Govt. of Belgium; BL/33/IN12). SJ, SCG, and OT acknowledge the financial support received from the BRICS grant under the project DST/IC/BRICS/2023/5. BN and RS acknowledge funding from the UNESCO-TWAS programme, "Seed Grant for African Principal Investigators" financed by the German Ministry of Education and Research (BMBF).  We acknowledge the use of SIMBAD and NASA’s ADS.  OV acknowledges funding from the Research Foundation Flanders (FWO), grant 1173025N. This paper includes data collected by the {\it Kepler} mission available at Mikulski Archive for Space Telescopes (MAST) operated by NASA. We thank the anonymous reviewers for a careful reading of the manuscript and for many insightful comments and suggestions that have improved the paper.

\bibliographystyle{mnras}
\bibliography{ref}

 \appendix
\section{Extra figures from \ktwo\ and \tess\ data analysis}

\begin{figure*}
\begin{center}
\includegraphics[width=\textwidth]{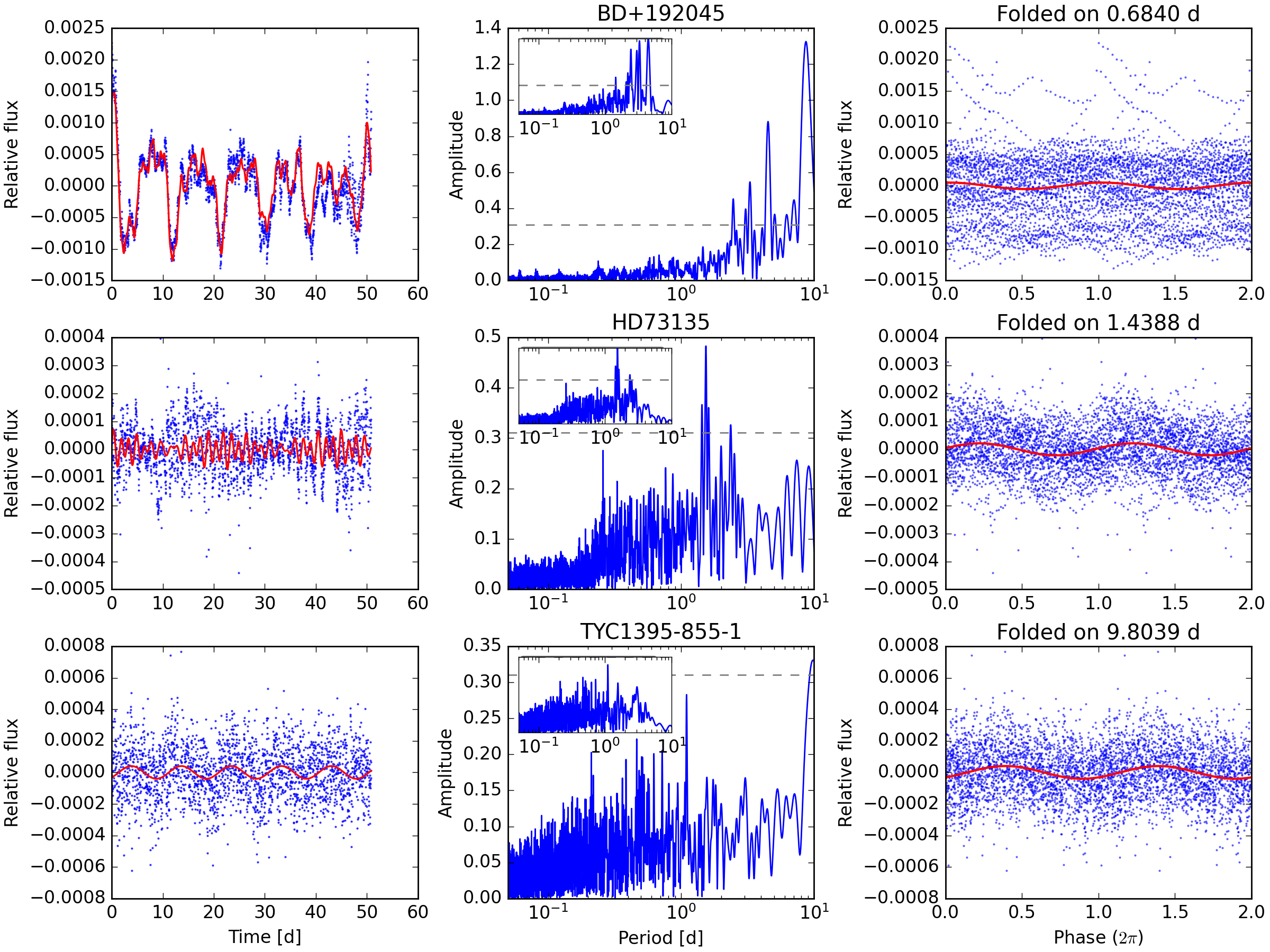}
\end{center}
\begin{center}
\caption{Same as Fig.\,\ref{K2_2015_Periodograms} but for long cadence  observations made during the 2018 K2 Campaign 18. Note that BD\,+19\degr\,2046 was not observed during this campaign.}
\label{K2_2018_Periodograms}
\end{center}
\end{figure*}

\begin{figure*}
\centering
 \includegraphics[width=\textwidth]{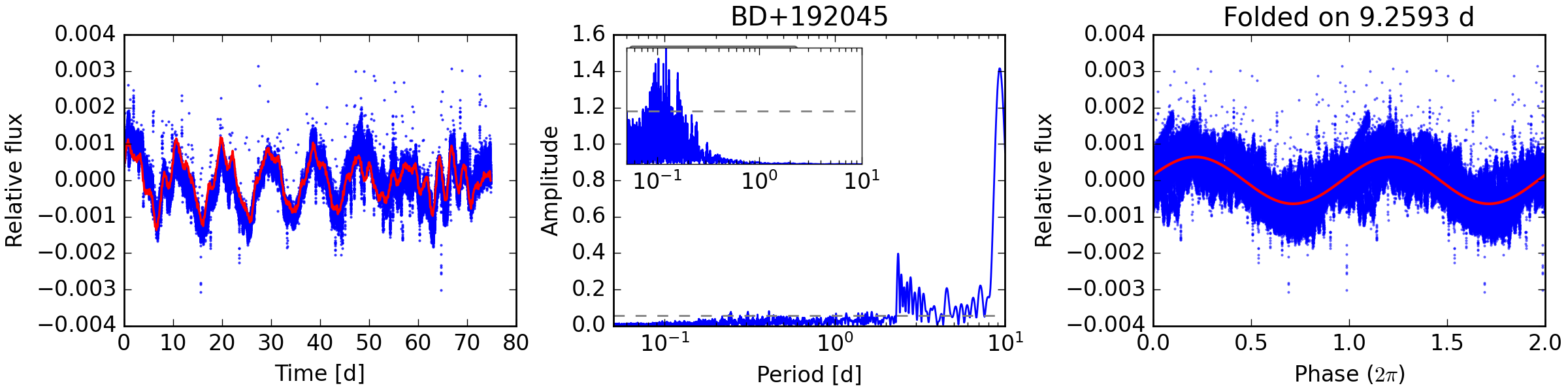}
\caption{Same as Fig.\,\ref{K2_2015_Periodograms} but for BD+19\degr\,2045 short cadence \ktwo\ data.}
   \label{BD+192045_sc}
\end{figure*}

\begin{figure*}
\centering
 \includegraphics[width=\textwidth]{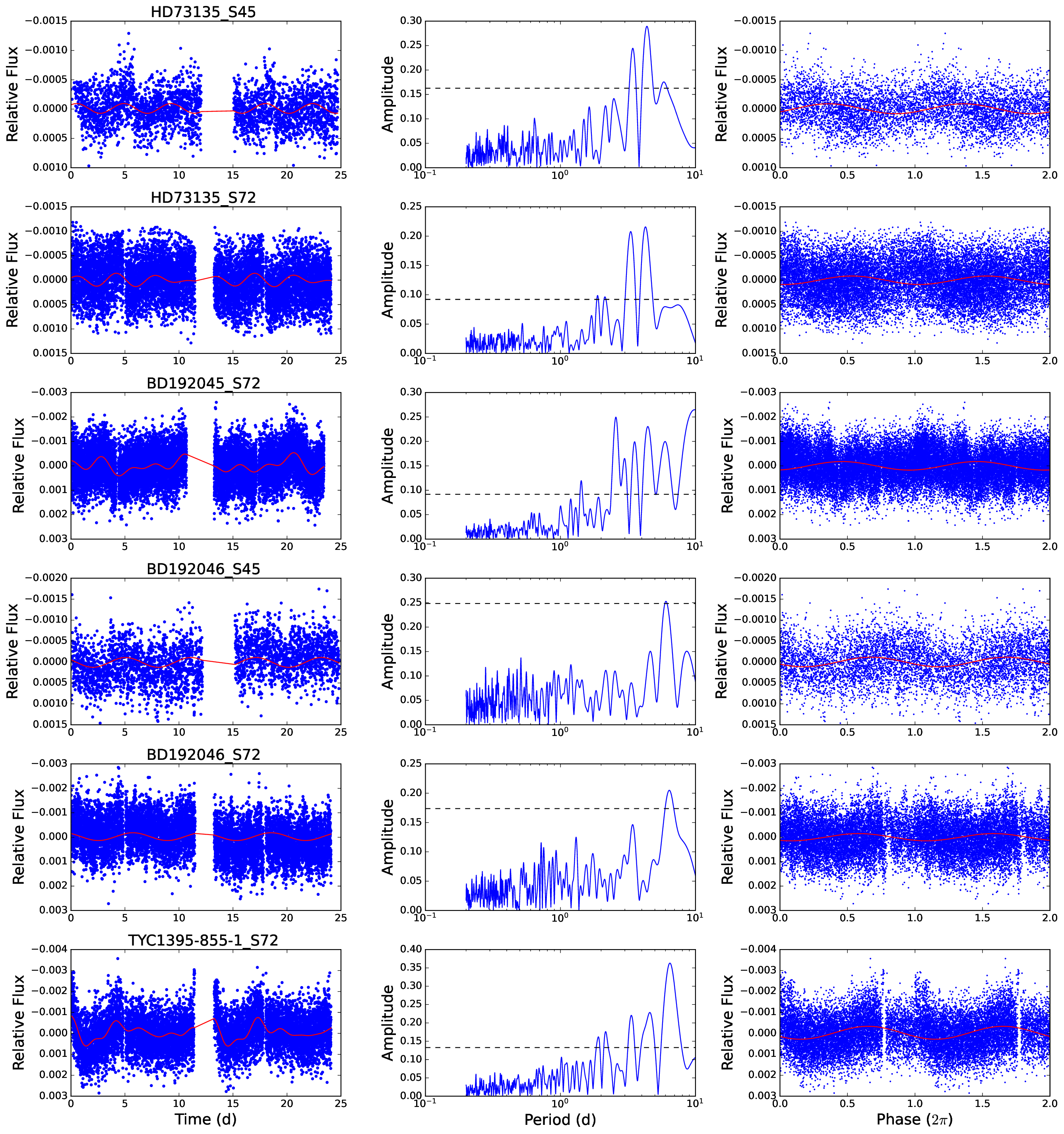}
\caption{Same as Fig.\,\ref{K2_2015_Periodograms} but for \tess\ data.}
   \label{fig:ft_tess}
\end{figure*}

\section{Time frequency analysis Figs.}

 \begin{figure*}
\centering
  \includegraphics[width=0.7\textwidth]{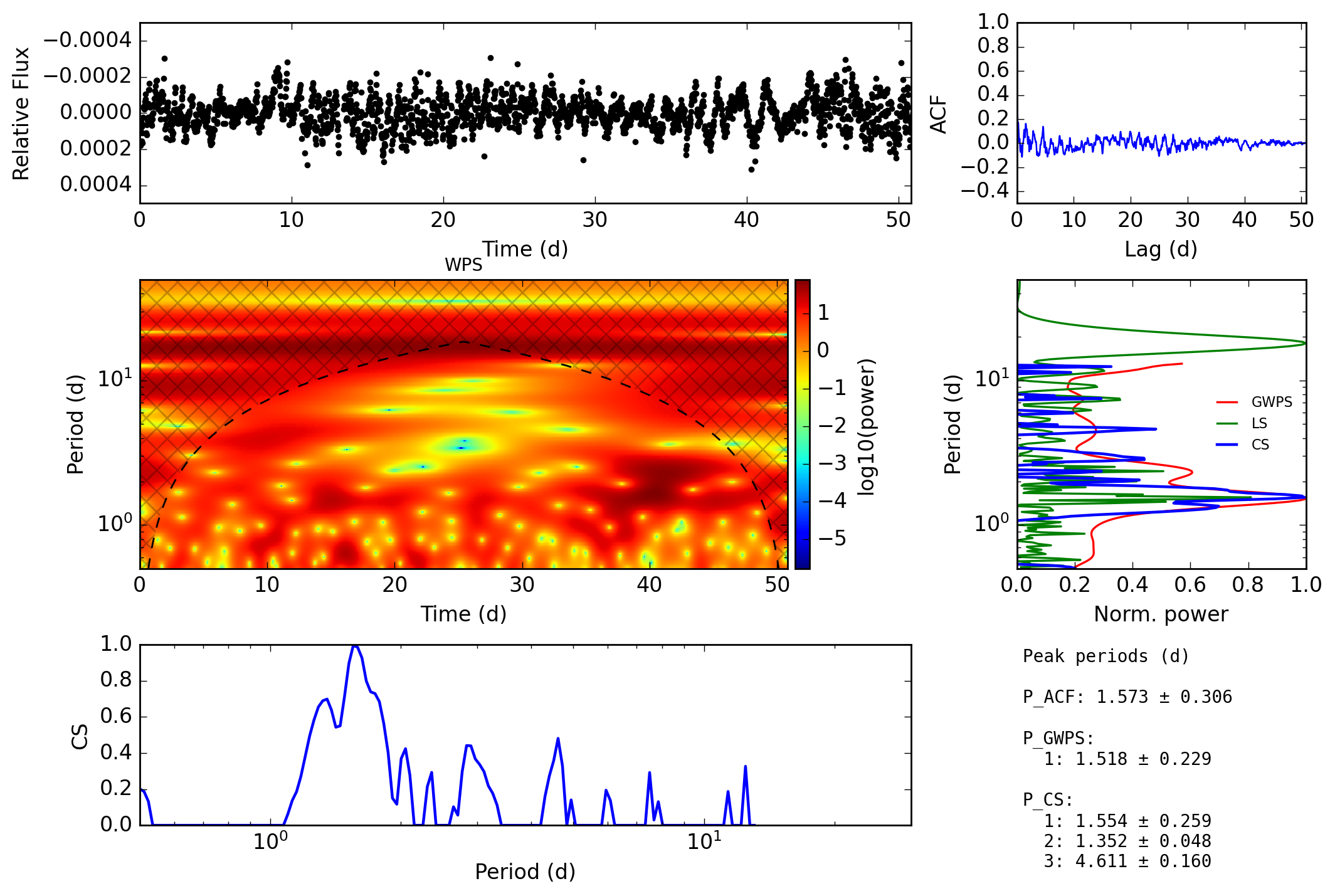}
  \caption{Same as for HD\,73135 in Fig.\,\ref{wavel2} but for campaign 18 long cadence mode.}
  \label{wavel3}
\end{figure*}

\begin{figure*}
\centering
  \includegraphics[width=0.7\textwidth]{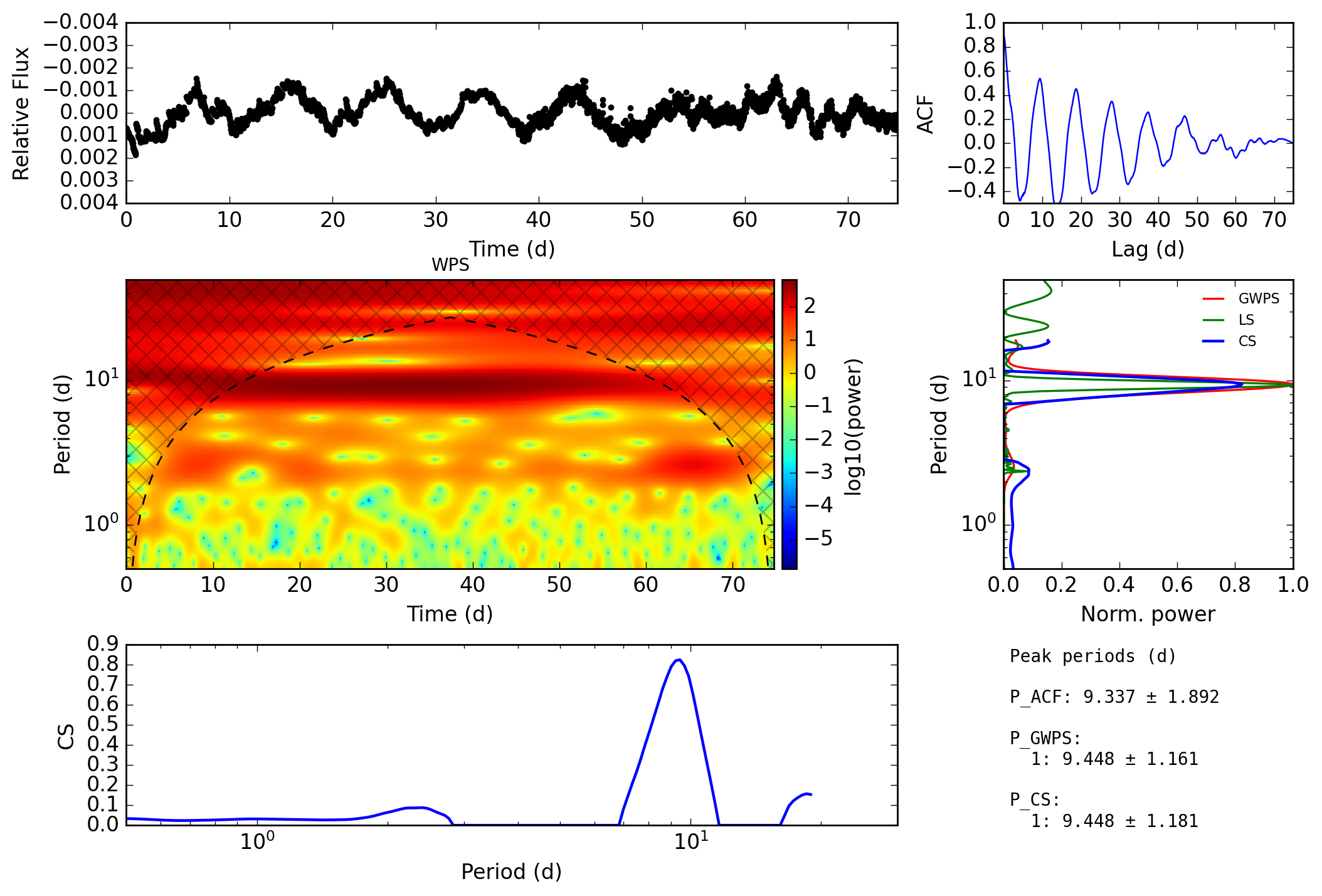}
  \caption{ Same as for HD\,73135 in Fig.\,\ref{wavel2} but for BD\,+19\degr\,2045 campaign 05 long cadence mode.}
  \label{wavel4}
\end{figure*}

\begin{figure*}
\centering
  \includegraphics[width=0.7\textwidth]{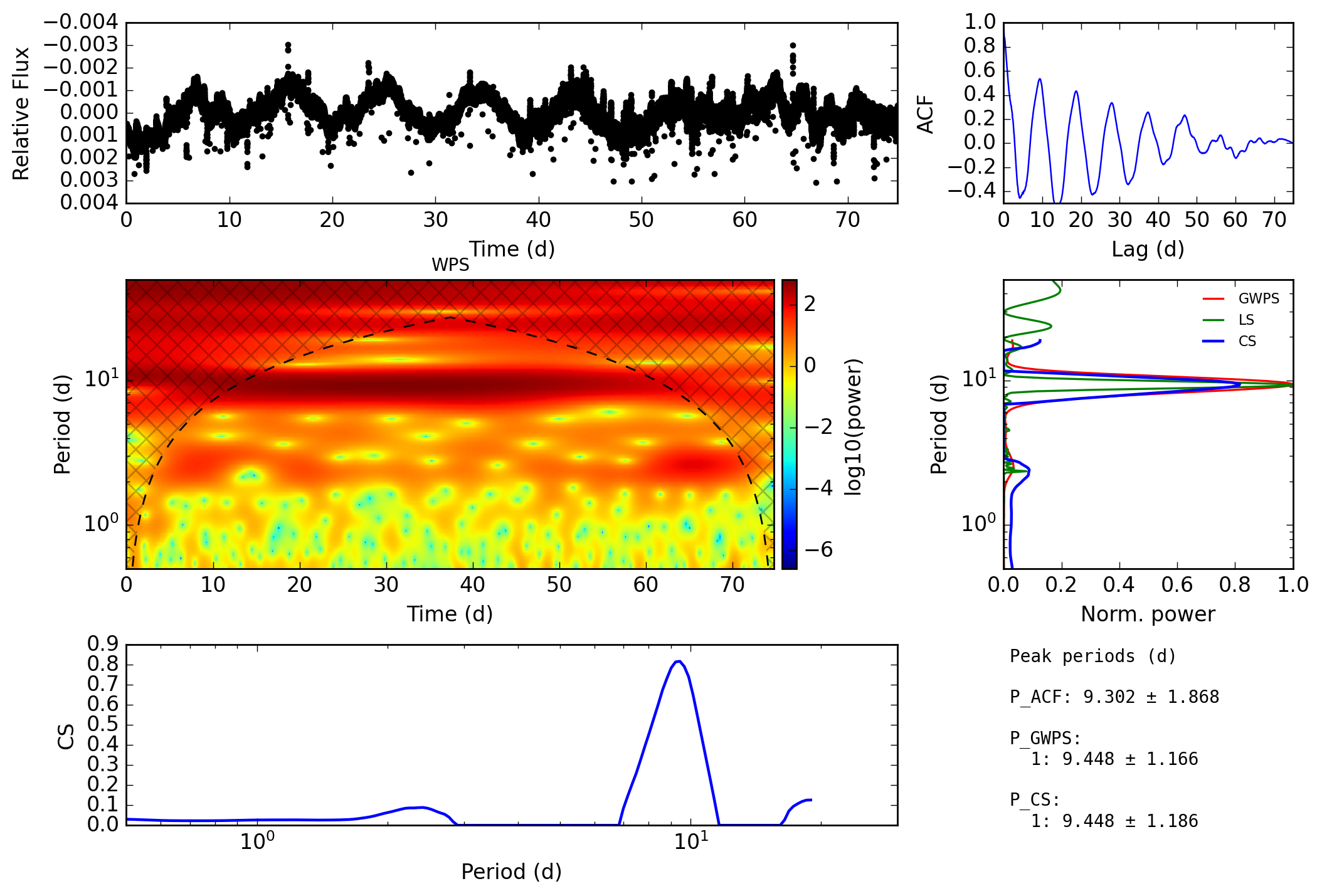}
  \caption{ Same as for HD\,73135 in Fig.\,\ref{wavel2} but for BD\,+19\textdegree\,2045 campaign 05 short cadence mode.}
  \label{wavel5}
\end{figure*}

\begin{figure*}
\centering
  \includegraphics[width=0.7\textwidth]{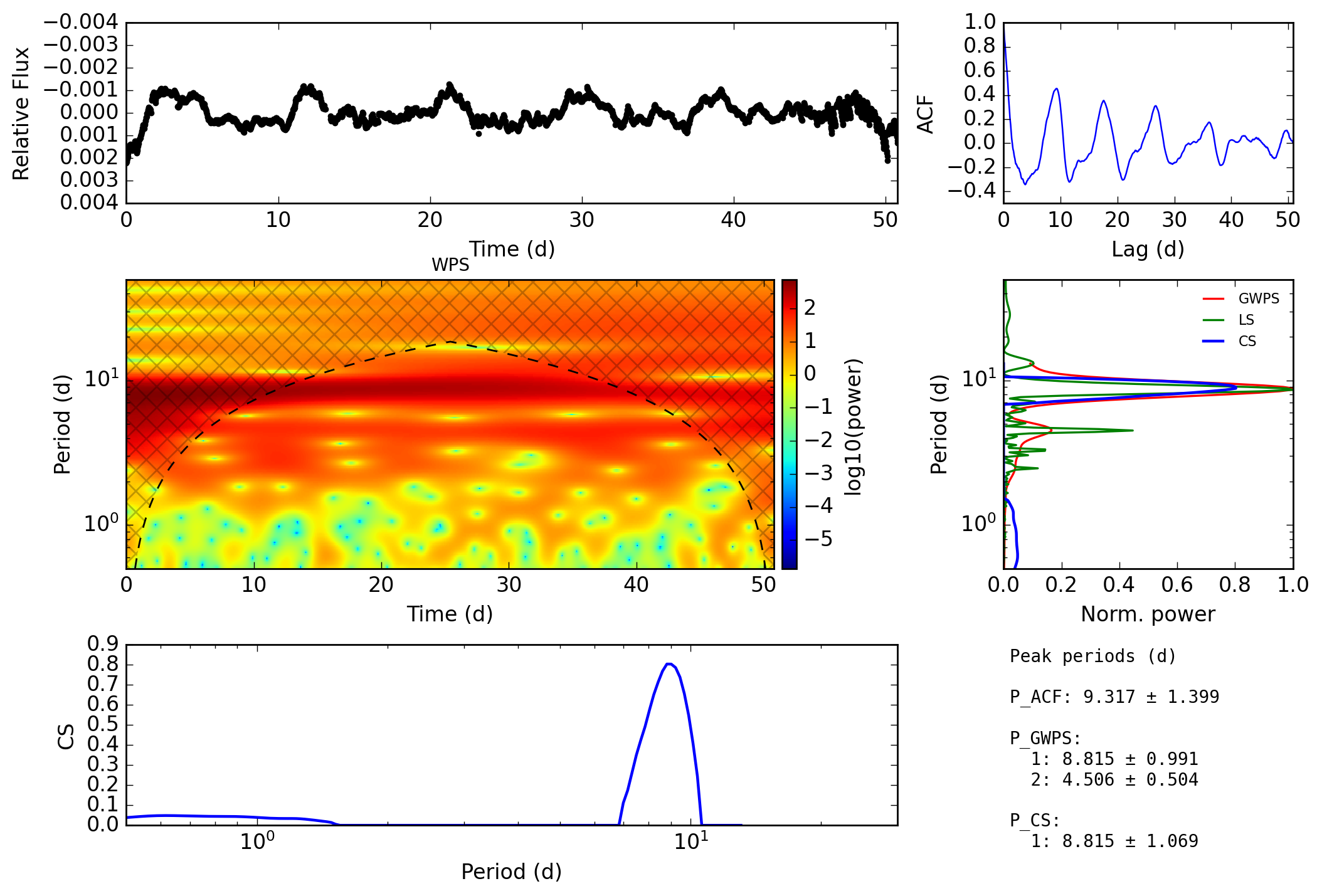}
  \caption{ Same as for HD\,73135 in Fig.\,\ref{wavel2} but for BD\,+19\textdegree\,2045 campaign 18 long cadence mode.}
  \label{wavel6}
\end{figure*}

\begin{figure*}
\centering
  \includegraphics[width=0.7\textwidth]{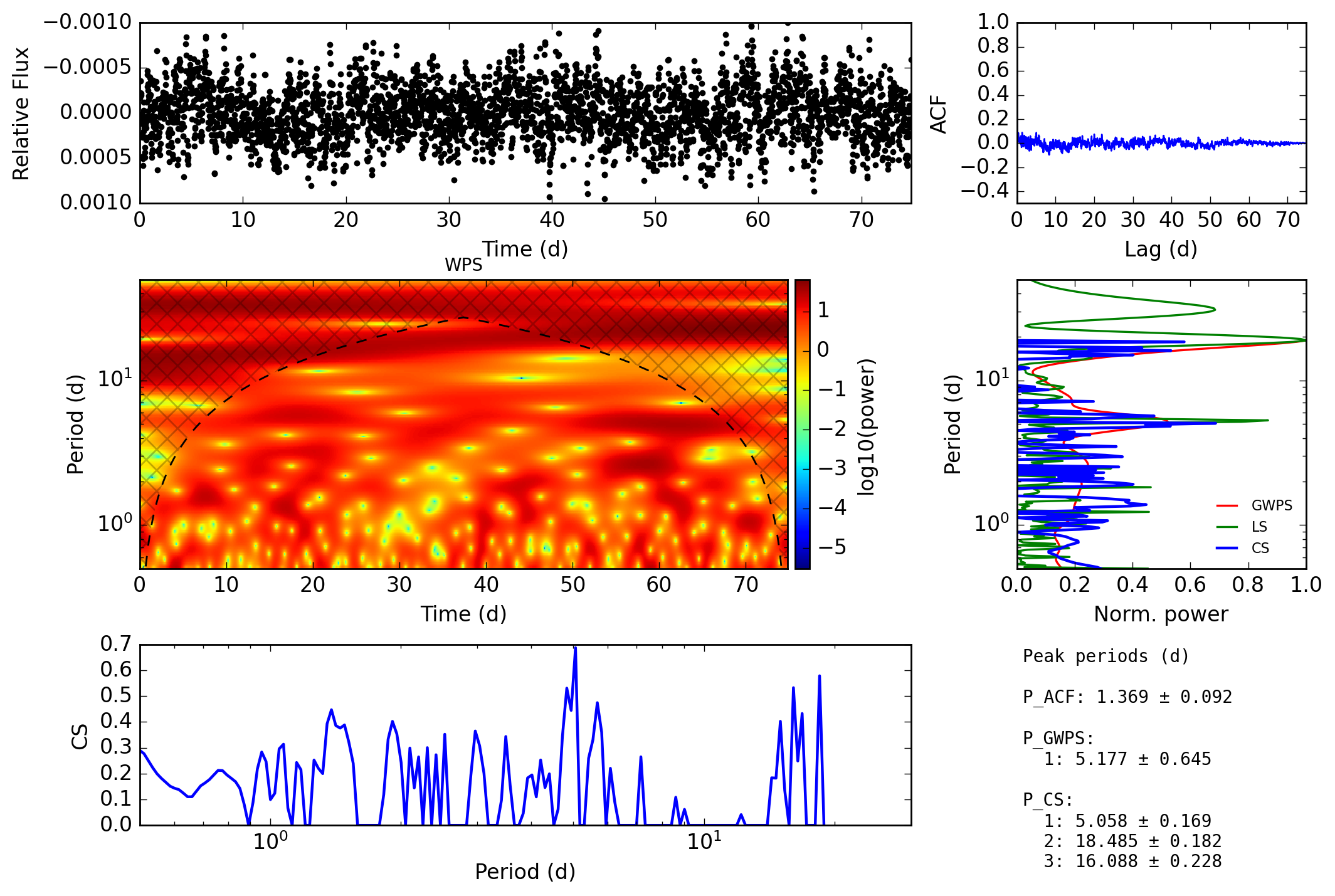}
  \caption{ Same as for HD\,73135 in Fig.\,\ref{wavel2} but for BD\,+19\degr\,2046 campaign 05.}
  \label{wavel7}
\end{figure*}

\begin{figure*}
\centering
  \includegraphics[width=0.7\textwidth]{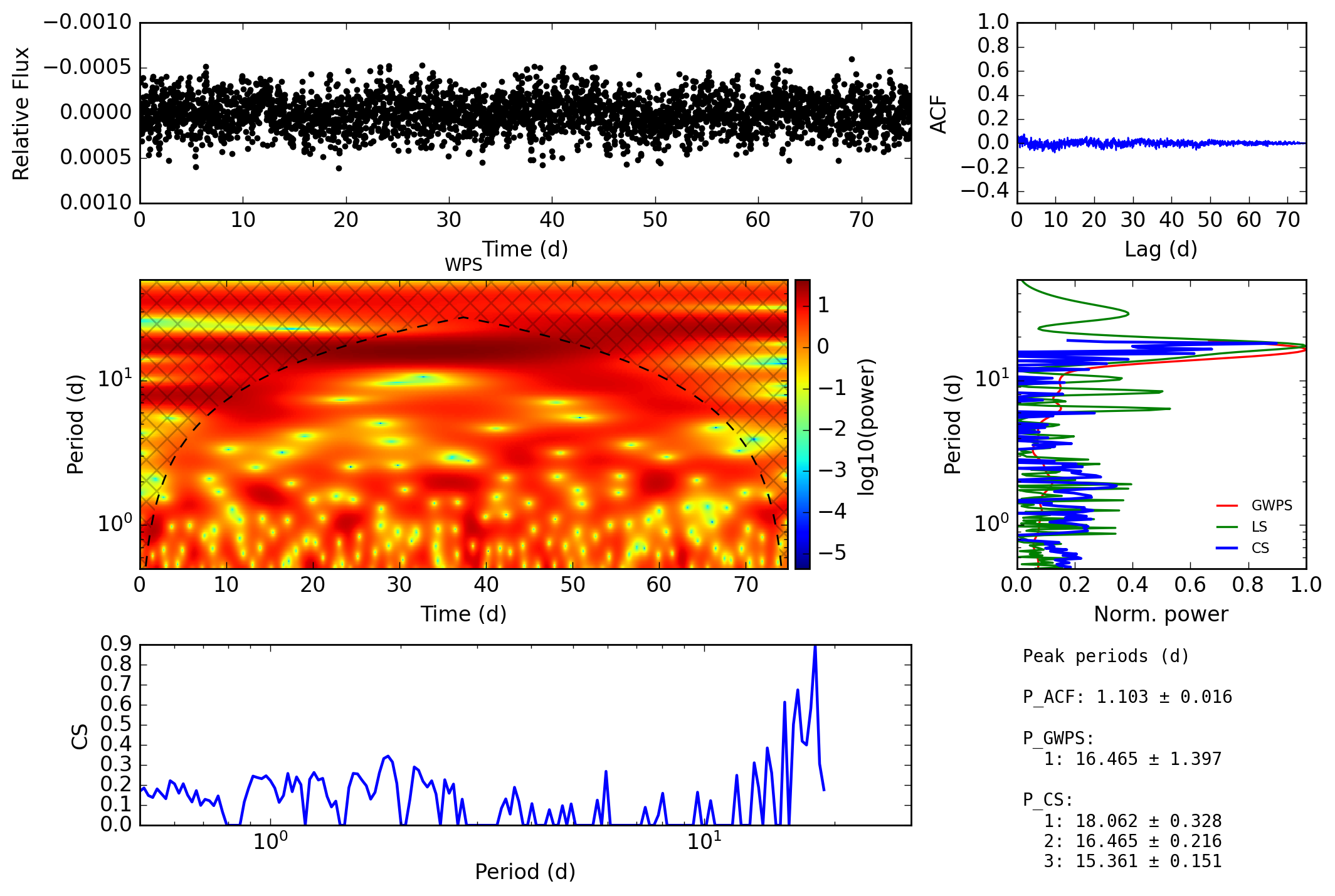}
  \caption{ Same as for HD\,73135 in Fig.\,\ref{wavel2} but for TYC\,1395-855-1 campaign 05.}
  \label{wavel8}
\end{figure*}

\begin{figure*}
\centering
  \includegraphics[width=0.7\textwidth]{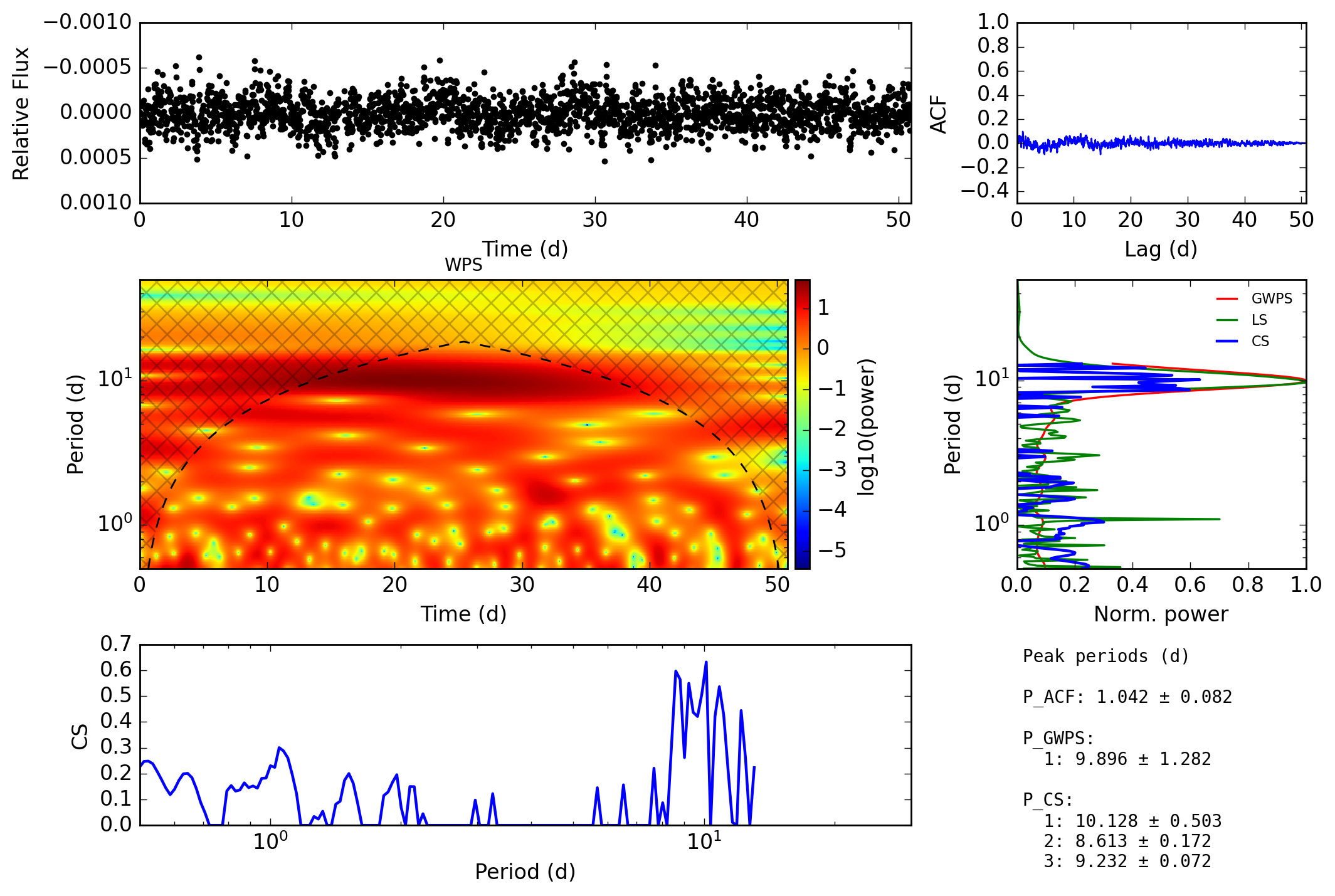}
  \caption{ Same as for HD\,73135 in Fig.\,\ref{wavel2} but for TYC\,1395-855-1 campaign 18.}
  \label{wavel9}
\end{figure*}

 \begin{figure*}
\centering
  \includegraphics[width=0.7\textwidth]{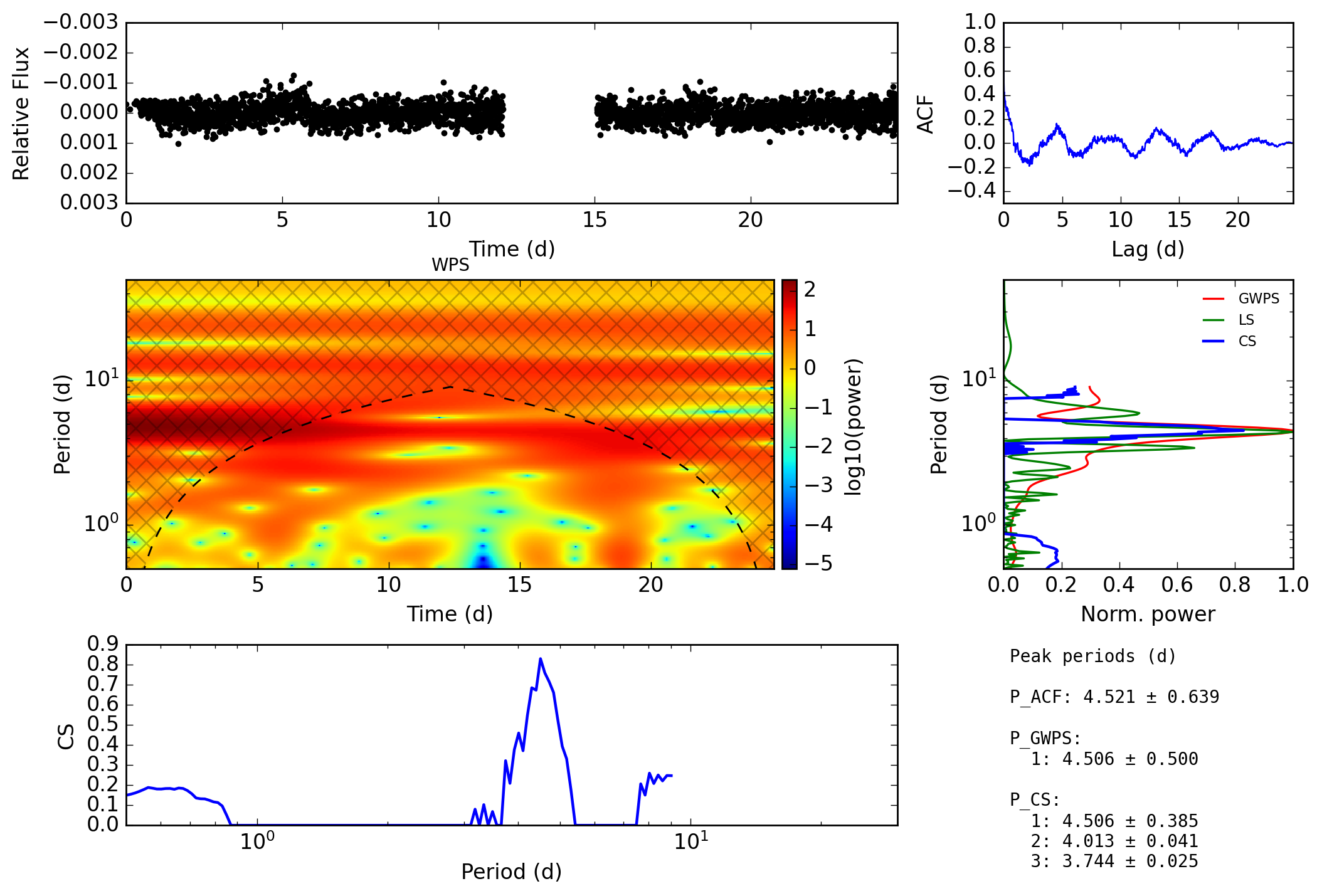}
  \caption{Same as for HD\,73135 in Fig.\,\ref{wavel2} but for \tess\ Sector 45.}
  \label{wavel10}
\end{figure*}

\begin{figure*}
\centering
  \includegraphics[width=0.7\textwidth]{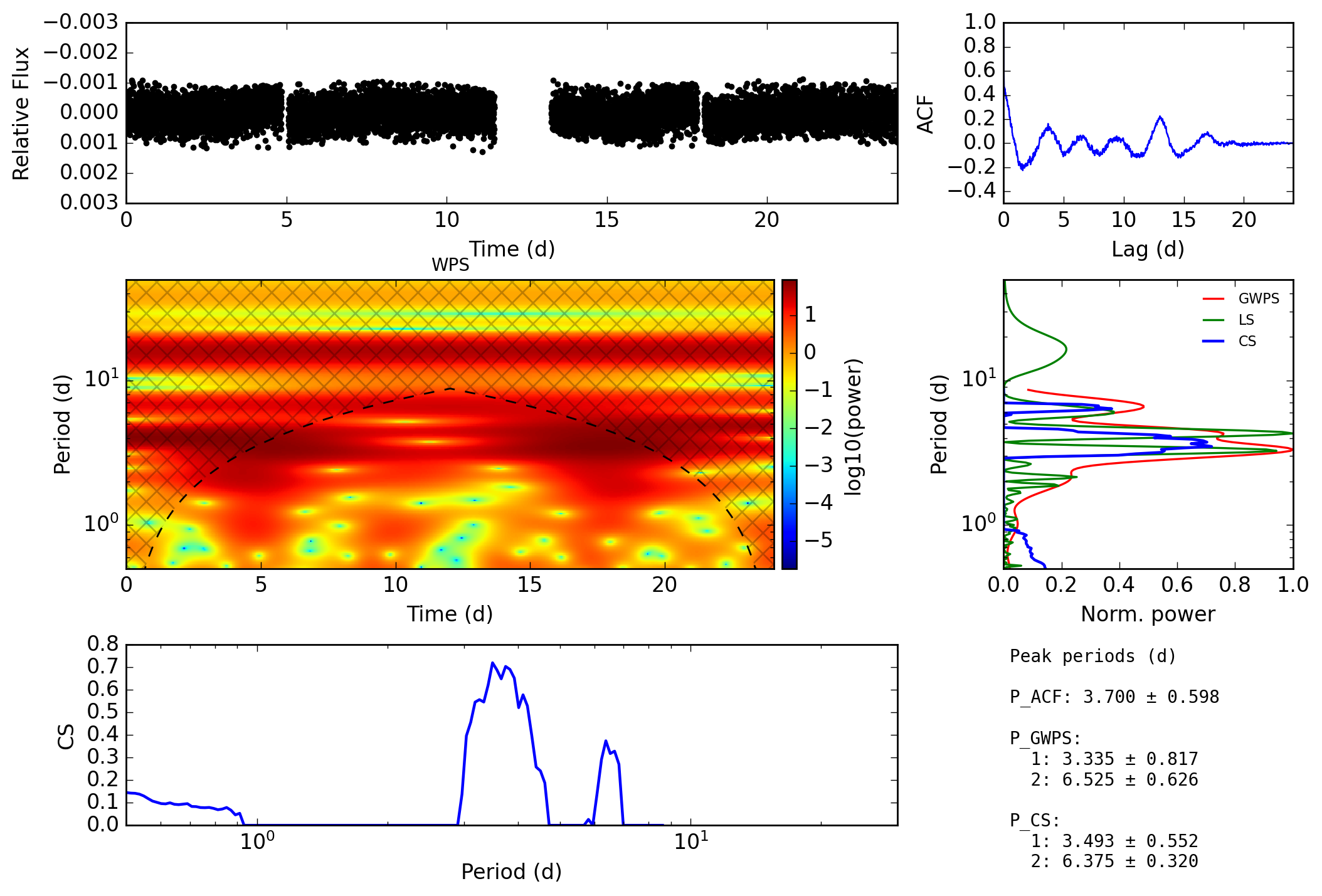}
  \caption{ Same as for HD\,73135 in Fig.\,\ref{wavel2} but for \tess\ Sector 72.}
  \label{wavel11}
\end{figure*}

\begin{figure*}
\centering
  \includegraphics[width=0.7\textwidth]{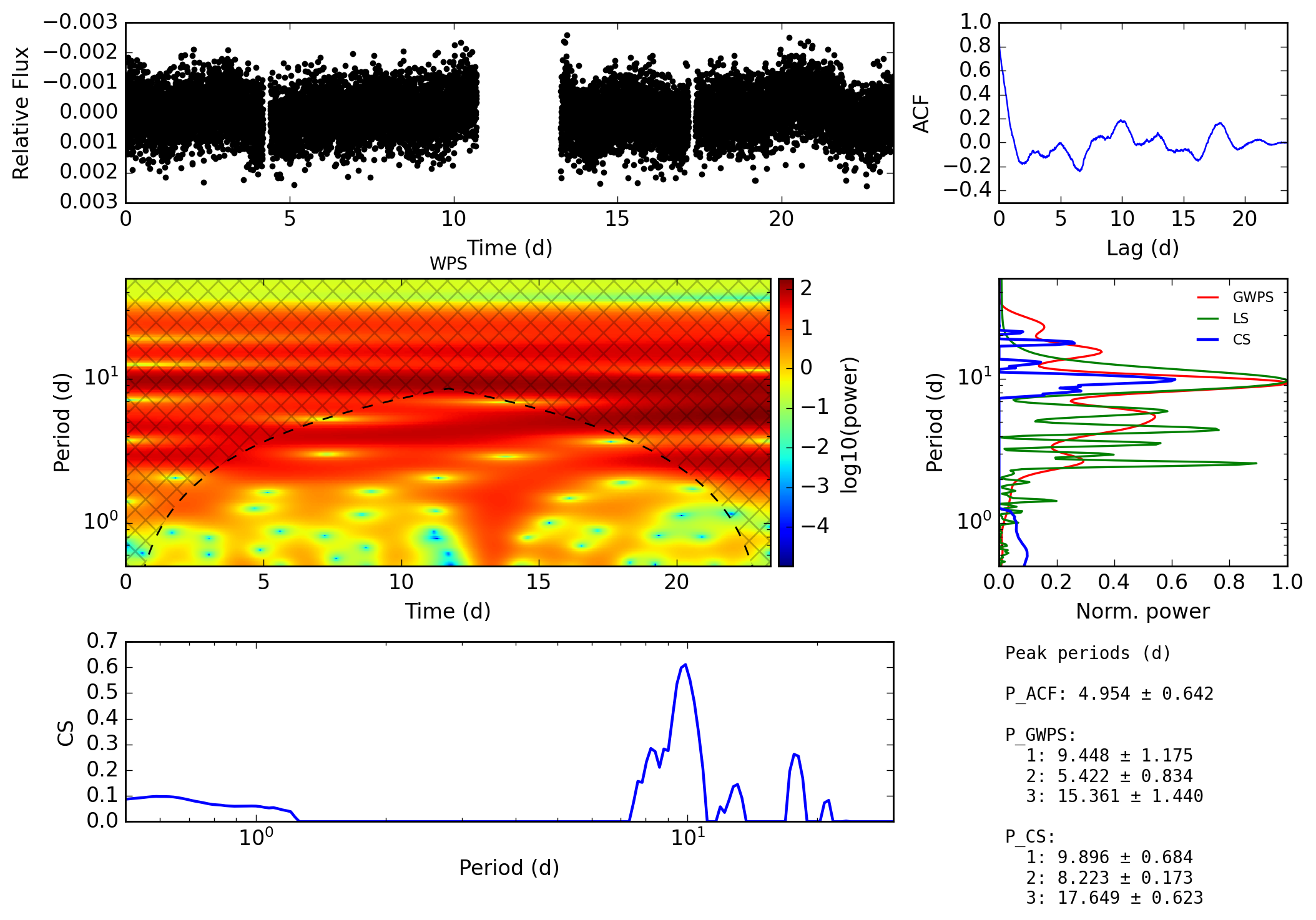}
  \caption{ Same as for HD\,73135 in Fig.\,\ref{wavel2} but for BD\,+19\textdegree\,2045 \tess\ Sector 72.}
  \label{wavel12}
\end{figure*}

\begin{figure*}
\centering
  \includegraphics[width=0.7\textwidth]{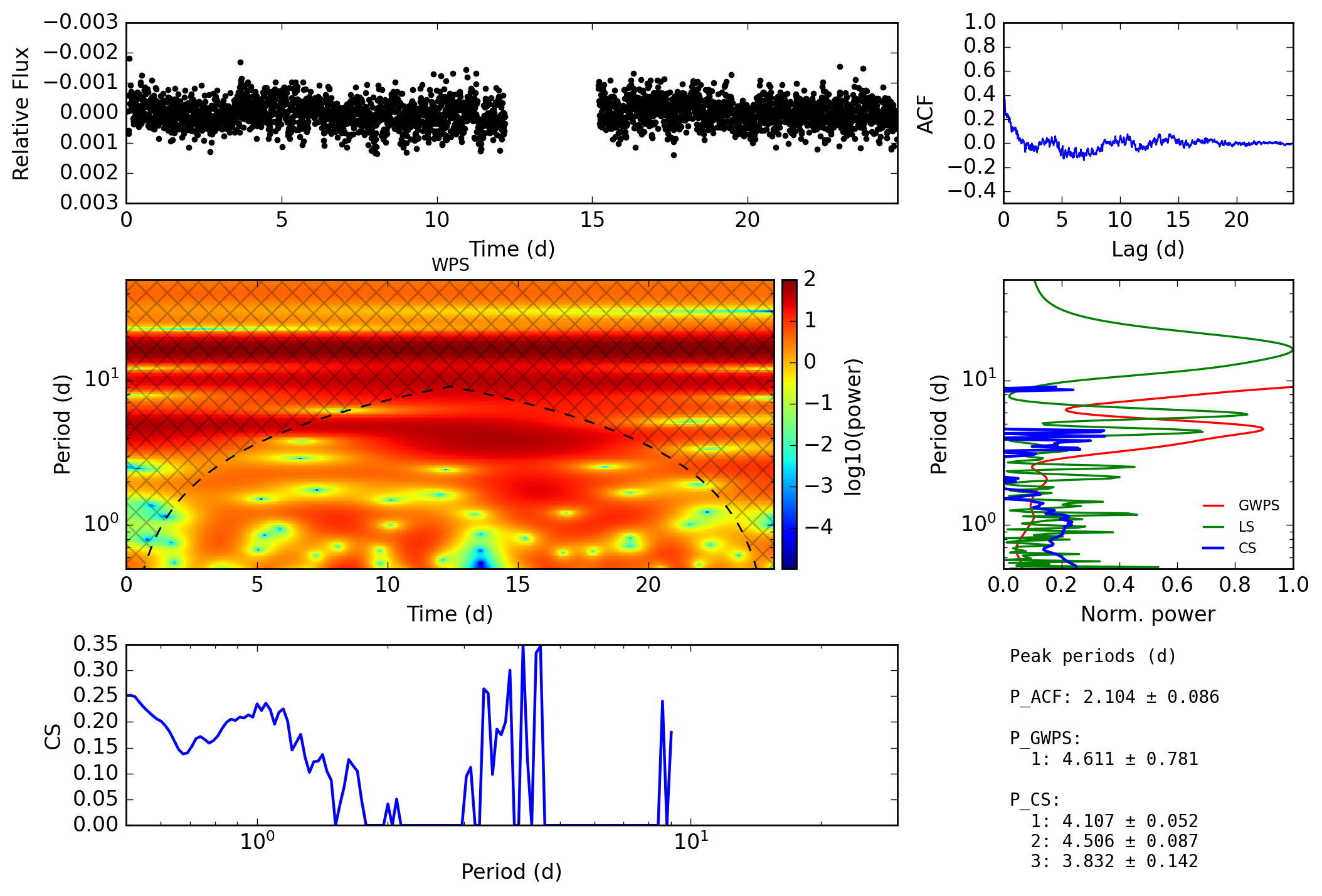}
  \caption{ Same as for HD\,73135 in Fig.\,\ref{wavel2} but for BD\,+19\textdegree\,2046 \tess\ Sector 45.}
  \label{wavel13}
\end{figure*}

\begin{figure*}
\centering
  \includegraphics[width=0.7\textwidth]{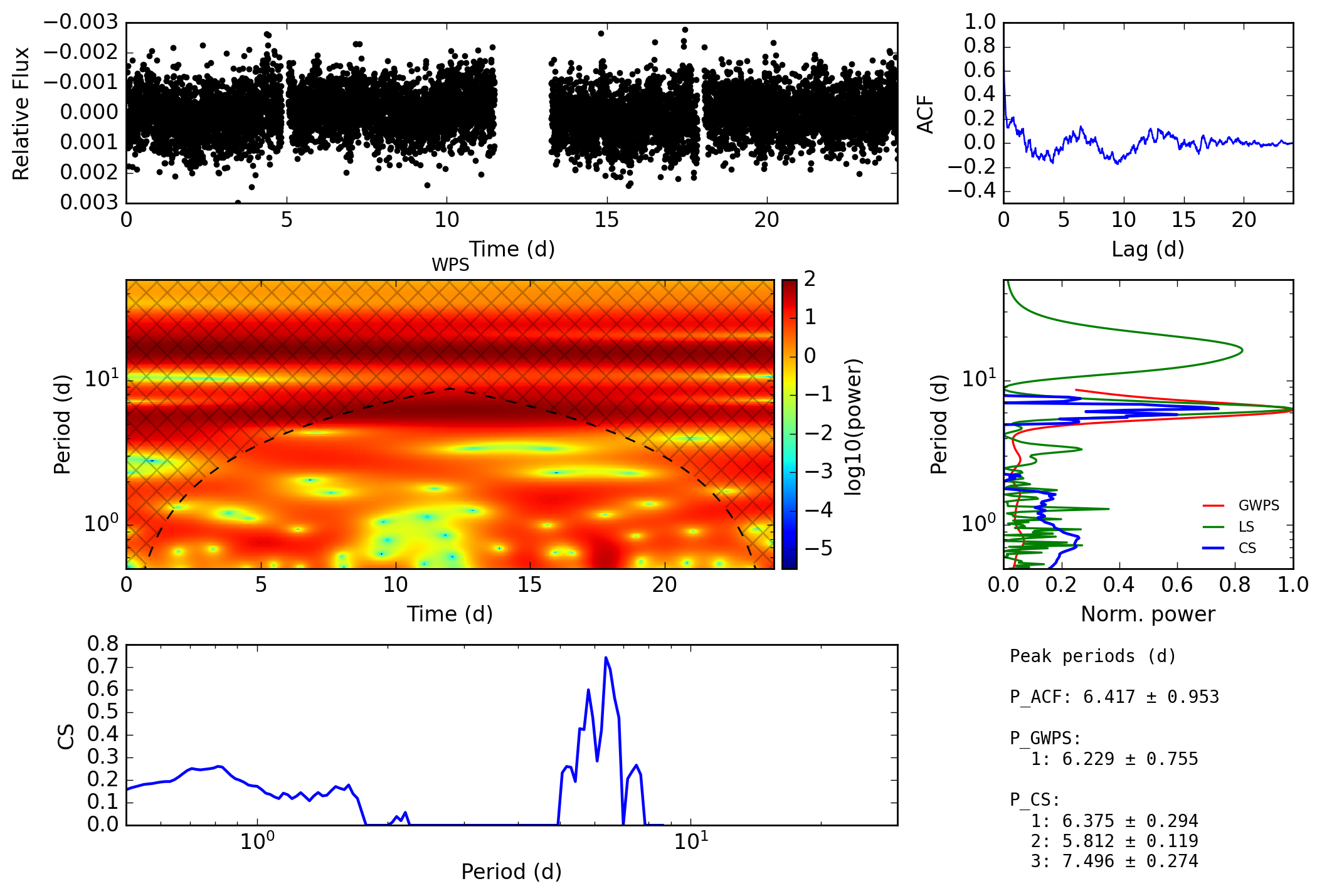}
  \caption{ Same as for HD\,73135 in Fig.\,\ref{wavel2} but for BD\,+19\degr\,2046 \tess\ Sector 72.}
  \label{wavel14}
\end{figure*}

\begin{figure*}
\centering
  \includegraphics[width=0.7\textwidth]{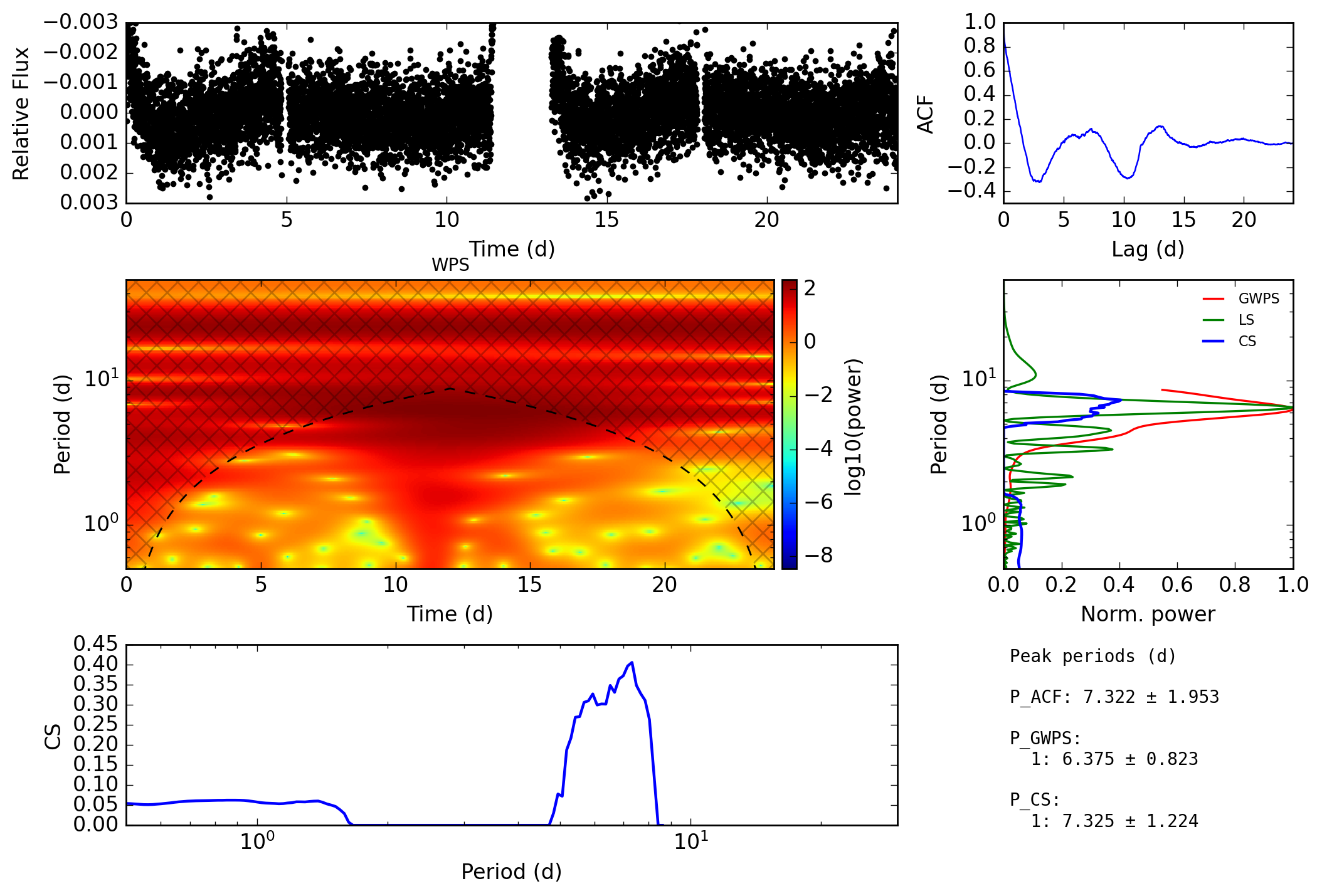}
  \caption{ Same as for HD\,73135 in Fig.\,\ref{wavel2} but for TYC\,1395-855-1 \tess\ Sector 72.}
  \label{wavel15}
\end{figure*}

 \bsp	
 \label{lastpage}
 \end{document}